\documentstyle[aps,epsf,prd,floats,cite,preprint]{revtex}

\def\xslash#1{{\rlap{$#1$}/}}
\def\Dsl{\hbox{/\kern-.6000em D}} 

\def\dsl{\,\raise.15ex\hbox{/}\mkern-13.5mu D} 
\def\bsigma{\mbox{\boldmath $\sigma$}}
\def\lqcd{\Lambda_{\rm QCD}}
\def\ms{$\overline{\rm MS}$}
\def\psip#1{\psi_{\mathbf{#1}}}
\def\chip#1{\chi_{\mathbf{#1}}}
\def\bsigma{\mbox{\boldmath $\sigma$}}
\def\muus{\mu_{\rm U}}
\def\mus{\mu_{\rm S}}
\def\abs#1{\left| #1 \right|}
\def\ltap{\ \raise.3ex\hbox{$<$\kern-.75em\lower1ex\hbox{$\sim$}}\ }
\def\gtap{\ \raise.3ex\hbox{$>$\kern-.75em\lower1ex\hbox{$\sim$}}\ }
\def\OMIT#1{}

\def\msb{{\overline{\rm MS}}}
\def\CL{{\cal L}}

\begin{document}

\tighten

\preprint{\vbox{ \hbox{UCSD/PTH 99--11} \hbox{UTPT-99-15} \hbox{CMU-99-11}}}
\title{Renormalization group scaling in nonrelativistic QCD} 
\author{Michael E.~Luke,${}^a$ Aneesh
  V.~Manohar,${}^b$ and Ira Z. Rothstein${}^c$}
\address{${}^a$ Department of Physics, University of Toronto,\\
  60 St.\ George Street, Toronto, Ontario, Canada M5S 1A7}
\address{${}^b$ Department of Physics, University of California at 
San Diego,\\
  9500 Gilman Drive, La Jolla, CA 92093-0319}
\address{${}^c$ Department of Physics, Carnegie Mellon University,\\
  Pittsburgh, PA 15213} \date{September 1999} \maketitle
\begin{abstract}
We discuss the matching conditions and renormalization group evolution of
non-relativistic QCD. A variant of the conventional $\msb$ scheme is proposed
in which a  subtraction velocity $\nu$ is used rather than a subtraction scale
$\mu$. We derive a novel renormalization group equation in velocity space which
can be used to sum logarithms of $v$ in the effective theory. We apply our
method to several examples. In particular we show that our formulation 
correctly reproduces the two-loop anomalous dimension of the heavy quark
production current near threshold.
\end{abstract}

\section{Introduction}

The dynamics of almost on-shell heavy quarks $Q$ with mass $m$ much greater
than the QCD scale $\lqcd$ can be computed in a systematic expansion in terms
of several small parameters. In the single heavy quark sector, the dynamics is
described by heavy quark effective theory (HQET), which has an expansion in
powers of $\alpha_s(m)$ and $\lqcd/m$~\cite{Book}. HQET can be used to compute
properties of hadrons such as the $\bar B$ and $D$ mesons containing a single
$b$ or $c$ quark. The dynamics in the quark-antiquark sector is far more
complicated than in the single quark sector. At low momentum transfer the $Q
\bar Q$ pair can form non-relativistic Coulomb-like bound states, which are the
$J/\psi$ and $\Upsilon$ for the $\bar c c$ and $\bar bb$ sectors,
respectively.  It should be possible to describe the dynamics of
nonrelativistic heavy quarks using a nonrelativistic effective field theory for
QCD. A formulation of this effective theory, called NRQCD (nonrelativistic
QCD), has been proposed by Bodwin, Braaten, and Lepage (BBL)~\cite{BBL}. The
analogous theory for electromagnetism, NRQED, was developed earlier by Caswell
and Lepage~\cite{Caswell:1986}.

Constructing NRQCD has proven to be more difficult than HQET, the complication
being that there are many scales involved. In HQET, the only two important
scales are the quark mass $m$ and $\lqcd$.  In NRQCD there two other important
scales, $m v$ and $m v^2$, the momentum and energy of the quarks (where $v$ is
the typical quark velocity).  Momentum regions with (energy, momentum) of order
$(m,m)$, $(mv, mv)$, $(mv^2, mv)$ and $(mv^2, mv^2)$ are referred to in the
literature as hard, soft, potential and ultrasoft,
respectively~\cite{Beneke:1997}.  The effective field theory must be able to
correctly reproduce phenomena in all of these regions.

The simplest approach to NRQCD uses a momentum space cutoff to regulate the
loop integrals.  This has the advantage that the physics below the cutoff
$\Lambda$ is automatically correctly taken into account.  However, the
usefulness of this  approach for computations is limited since cutoffs break
gauge invariance.  Furthermore, the theory does not have manifest power
counting---loop graphs mix powers of $v$.  If a mass independent subtraction
scheme such as \ms\ is applied to the BBL Lagrangian, the $v$ expansion breaks
down due to unphysical poles introduced by the nonrelativistic approximation. 
There have been many approaches advocated to remedy this situation.

In Ref.~\cite{Labelle:1998}, it was shown that it was more useful to formulate
NRQCD as a theory in which ultrasoft modes couple  via the multipole expansion.
A velocity power counting rule for bound states in nonrelativistic effective
field theories was formulated in Ref.~\cite{Luke:1997}. The leading order term
in the effective Lagrangian reproduced the form of the propagator in the
potential regime. To recover the poles in the gluon propagator that correspond
to gluon radiation, the gluon propagator $1/(v^2 (k^0)^2-\mathbf{k}^2)$ had to
include subleading terms in $v$, which caused problems with the naive velocity
power counting rules. In Ref.~\cite{Manohar:1997}, it was pointed out that the
usual matching onto NRQCD violated $v$ power counting if the \ms\ scheme was
used, and it was shown that the problem could be fixed in the single
heavy-quark sector by using the same matching conditions as for HQET. In
Ref.~\cite{Grinstein:1998}, it was  demonstrated that the multipole expansion
is the appropriate generalization of ~\cite{Manohar:1997}  to the two quark
sector. In Ref.~\cite{Luke:1998}, an effective theory was formulated using two
different fields for the potential and radiation gluons. A problem which arose
in this formulation, however, was that it neglected soft gluon modes, which are
responsible for the running of the coupling below $\mu=m$.

In the threshold expansion \cite{Beneke:1997}, the results of NRQCD are
obtained directly from QCD by expanding graphs about the relevant kinematic
regimes (hard, soft, potential and ultrasoft).  This technique has recently
been used to extract the two-loop corrections to top-antitop production near
threshold with comparative ease
\cite{Hoang:1997a,Hoang:1997b,Czarnecki:1998,Beneke:1998}.  However, it is less
simple to perform renormalization-group improved calculations in this
formulation than in a true effective field theory (our results in this paper
will disagree with the RGE analysis presented in \cite{Beneke:1999}.)  The
threshold expansion was written as an effective theory by Griesshammer
\cite{Griesshammer:1998}.

In the approach advocated by Pineda and Soto\cite{Pineda:1997a,Pineda:1997b,
Pineda:1998a,Brambilla:1999b} the matching onto the effective field theory
occurs at two stages. Matching between QCD and NRQCD occurs at the scale
$\mu=m$, while at $\mu$ of order the inverse separation between the heavy
quarks NRQCD is matched onto a new effective theory which the authors call
pNRQCD (p for potential). In particular, Pineda and Soto argue that the
matching between QCD and NRQCD should contain only the hard part of loop
integrals, and should be performed using HQET Feynman rules. By performing the
matching exactly at threshold, the Coulomb singularity is regulated by
dimensional regularization, so the one-loop matching conditions are well
defined.  Furthermore, the treatment of soft modes is particularly simple in
this approach, as they just correspond to the running in the theory between $m$
and $mv$.

We argue in this paper, however, that the problem with  this approach is that
HQET Feynman rules do not correctly treat the momentum region between $m$ and
$mv$.  In particular, in \cite{Pineda:1998a} it is argued that the anomalous
dimension for the electromagnetic current for heavy quark production vanishes.
While this is true at one loop, at two loops the current has a nonzero
anomalous dimension~\cite{Hoang:1997a,Hoang:1997b} which HQET Feynman rules
cannot reproduce.

In this paper, we construct an effective theory for NRQCD which has a
consistent $v$ expansion when loops are evaluated in the \ms\ scheme, and which
correctly reproduces the two loop anomalous dimension of the heavy quark
production current.   The Lagrangian we use  is similar to that of
\cite{Griesshammer:1998}, however, we do not have to introduce as many extra
fields (such as soft quarks) as in that formulation.  Unlike the pNRQCD
approach, we argue that the correct matching scale onto the effective
Lagrangian (similar to that of pNRQCD) is $\mu=m$, not $\mu=mv$.  The added
complication which then arises is that soft modes must explicitly be taken into
account between $\mu=m$ and $\mu=mv$, in order to obtain the correct running of
the potential. 

We also introduce a novel renormalization group (VRG) equation in velocity
space that is used to sum logarithms of $v$ in the effective theory. The VRG
represents the invariance of the theory under changes in the subtraction
velocity $\nu$. The formulation of NRQCD presented in this paper allows one to
include the effects of the running coupling constant in the quark potential by
using the velocity renormalization group equations, and to simultaneously sum
soft and ultrasoft logarithms.

In Sec.~\ref{sec:scales}, we discuss some general aspects of the problem, and
in Sec.~\ref{sec:eft} we introduce the fields required in the effective field
theory and discuss power counting and loop graphs in NRQCD.   The VRG is
introduced in Sec.~\ref{sec:VRG}, while in Sec.~\ref{sec:example} we illustrate
the formalism with some examples.  In particular, we show that we correctly
reproduce the two-loop anomalous dimension of the heavy quark production
current.  We defer the complete RGE analysis of heavy quark production to a
future paper.

\section{{\large {${\bf{p^2}}/{2m}$}} or no {\large{${\bf{p^2}}/{2m}$}} }
\label{sec:scales}

Consider pair production of a $\bar Q Q$ pair near threshold by a virtual
photon.  We are interested in the threshold region, where the fermions are
nonrelativistic, so that $v\ll 1$, where
\begin{equation}\label{5}
v= \sqrt{1 - {4 m^2 \over s}}
\end{equation}
is the velocity of the two final state fermions (ignoring for the moment
complications from confinement effects in QCD).  The electromagnetic current in
the full theory matches to
\begin{eqnarray}\label{eff-1}
J^i &=& \psi^\dagger \sigma^i \left[\chi^\dagger\right]^T C(\mu),\nonumber\\
C(\mu)&=& 1 +c_1(\mu)
{{\alpha_s}\over{\pi}} + c_2(\mu) \left({\alpha_s\over{\pi}}\right)^2+
\dots ,
\end{eqnarray}
where $\psi$ and $\chi$ annihilate quarks and antiquarks, respectively.
Ignoring for the moment non-perturbative effects, there are three relevant
scales in the process: the quark mass $m$, the quark three-momentum $p=mv$, and
the quark energy $E=mv^2/2$.  

An approach to the problem of unraveling these scales was developed in
Refs.\ \cite{Manohar:1997,Pineda:1997a,Pineda:1997b,
Pineda:1998a,Brambilla:1999b}. The authors argued that at the scale $m$, no
distinction need be made  between energy and momentum, since they are both $\ll
m$; it is only at the scale $mv$ that they are distinguished in the power
counting. The correct effective field theory was therefore argued to be
identical to HQET. The NRQCD and HQET descriptions differ in how they treat
these scales. In the HQET approach, the kinetic term in $\CL$ is taken as a
perturbation, while in the NRQCD approach, the kinetic term is resummed in the
propagator.  While this violates $m$ power counting, it was shown in
\cite{Luke:1998} that as long as the potential is taken to be instantaneous,
and real radiation is coupled via the multipole expansion, there is a
consistent counting in $v$.

At one loop, both approaches yield the correct result for the matching onto the
external current.  In the NRQCD approach the Coulomb singularity $\sim 1/v$ in
the $\bar Q Q$ production amplitude is reproduced by nonrelativistic fermions
undergoing instantaneous potential exchange, while in the HQET
approach the Coulomb singularity is regulated at threshold by dimensional
regularization. In the latter case the matching condition is given solely by the 
hard part of the graph, obtained by evaluating the full theory at threshold.  Both
approaches give the well-known result for the matching condition,
\begin{equation}\label{anom1}
c_1(\mu)=-2 C_F.
\end{equation}
At two loops, however, the approaches differ.  The two-loop matching onto NRQCD
was computed by Hoang for QED~\cite{Hoang:1997a,Hoang:1997b}, and the
computation has been recently extended to the non-Abelian case by Czarnecki and
Melnikov~\cite{Czarnecki:1998}, and by Beneke, Signer and
Smirnov~\cite{Beneke:1998}. These authors find
\begin{equation}\label{eff-2}
c_2(\mu)= \pi^2 C_F \left({1\over3}C_F + {1\over2}C_A\right)
\ln {m\over \mu}+\mbox{non-logarithmic\ terms}.
\end{equation}
The electromagnetic current has no anomalous dimension in the full theory,
which implies that in the effective theory, $C$ must have an anomalous
dimension at $\mu=m$,
\begin{equation}\label{anom2a}
\mu{d C(\mu)\over d\mu} = -\pi^2 C_F \left({1\over3}C_F + {1\over2}C_A\right)
{\alpha_s}^2.
\end{equation}

The anomalous dimension is of leading order in the $1/m$ expansion,
and it is straightforward to verify in either $A^0=0$
or Coulomb gauge that the leading order graphs do not give a
two-loop anomalous dimension in HQET.  

\begin{figure}
\epsfxsize=3cm
\hfil\epsfbox{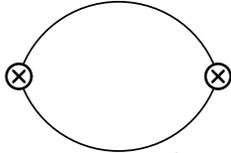}\hfill
\caption{Fermion-antifermion propagation graph. The two $\otimes$ create and
annihilate a fermion-antifermion pair.
\label{fig:pair}}
\end{figure}

The situation is rather different in NRQCD, which has a $v$ power counting
scheme. In this case, the $O(1)$ anomalous dimension arises at two loops due to
a $1/v^2$ enhancement of an $O(v^2)$ term in the potential: the $1/v^2$ Coulomb
enhancement is crucial to this result. We will compute the anomalous dimension
in Sec.~\ref{hoang}, and show that it correctly reproduces Eq.~(\ref{anom2a}).
The distinguishing feature between NRQCD and HQET is that HQET does not have
the Coulomb divergence. By evaluating one-loop graphs exactly at threshold one
avoids the problem of Coulomb divergences, and this procedure allows one to
compute the one-loop matching correction Eq.~(\ref{anom1}) using HQET. However,
in two-loop graphs the internal graph is not at threshold, and so is sensitive
to the Coulomb singularity.  The problem in the $m$ counting scheme seems to be
that unless the ${\bf p}^2 /2 m$ term is included in the leading order
propagator, the effective theory cannot correctly reproduce the propagation of
a fermion-antifermion pair, such as the graph in Fig.~\ref{fig:pair}, which
vanishes in dimensional regularization if HQET propagators are used.

Thus the effective field theory at $\mu=m$ must resum the ${\mathbf p}^2/2m$
term in the propagator to reproduce the infrared physics of full QCD, and to
correctly reproduce the two-loop anomalous dimension. Once the ${\mathbf
p}^2/2m$ term is included in the quark propagator, it is also necessary to
perform a multipole expansion and include a quark-antiquark
potential~\cite{Grinstein:1998}. The matching from QCD to an effective theory
with potentials is done at the scale $\mu=m$, so the potential in the effective
theory at $\mu=m$ depends on $\alpha_s(m)$.  Since the dominant momenta in the
static potential are of order $m v$, one might expect that the relevant
coupling is $\alpha_s(m v)$, and this is borne out by more detailed studies of
the quark static
potential~\cite{Appelquist:1977,Appelquist:1978,Fischler:1977}. One therefore
requires that the potential generated at $\mu=m$ must run in the effective
theory below $m$. This running can be implemented by the inclusion of soft
gluon modes, with $(E,{\mathbf p})\sim (mv,mv)$, the importance of which was
pointed out by Greisshammer~\cite{Griesshammer:1998}.  Soft gluon modes should
be integrated out of NRQCD, since they can never be produced on-shell;
nevertheless they must be included in the running between $m$ and $mv$.

\subsection{Possible Hierarchies}

In addition to the scale $m v$ and $m v^2$ the non-perturbative scale $\lqcd$
also plays an important role for real quarkonium. Though $\lqcd$ will not play
an important role for the analysis of this paper, some aspects of its power
counting are worth emphasizing.

\begin{figure}
  \epsfxsize=8cm \hfil\epsfbox{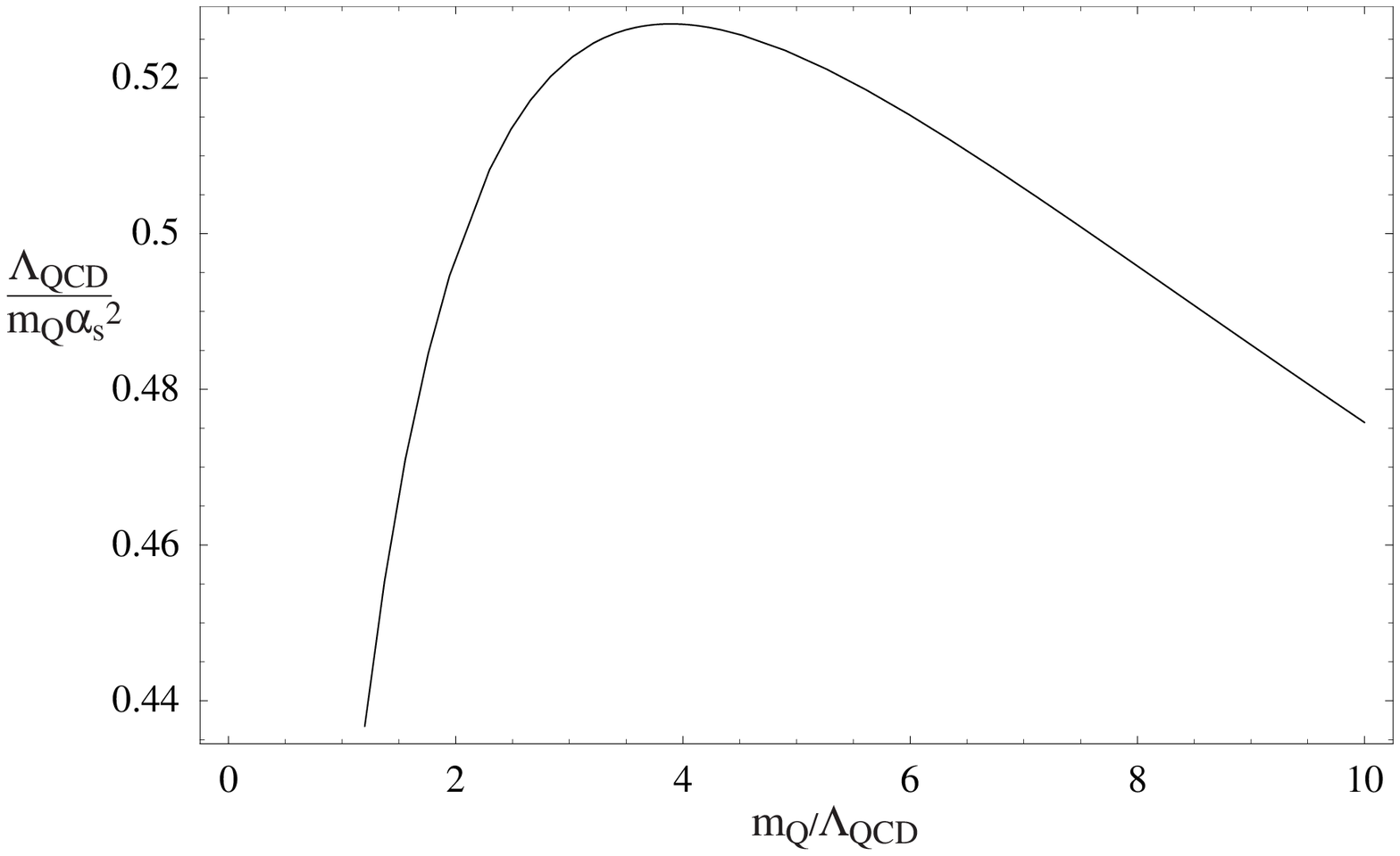}\hfill
\caption{Plot of $\lqcd/m\alpha_s^2(m v)$ as a function of $m/\lqcd$, for
$n_f=3$. \label{fig:3}}
\end{figure}

For very large $m$, or  equivalently, small $\alpha_s(m)$, one is in the regime
$\lqcd \ll m v^2 \ll m v \ll m$, since $\lqcd$ is formally smaller than any
power of $\alpha_s$.   These inequalities are only well satisfied for $t$
quarks; for charmonium and bottomium the situation is closer to  $\lqcd\sim
mv^2$ or $\lqcd\sim mv$, and non-perturbative effects become important.  Of
course, the apparent independence of $m v$, $m v^2$ and $\lqcd$ for a Coulomb
system is illusory. The velocity $v$ in a Coulomb bound state is given by
solving $v=\alpha_s(m v)$,
\begin{equation}\label{7a}
v= {4 \pi \over b_0 \ln \left(m^2 v^2/\lqcd^2\right)},
\end{equation}
where
\begin{equation}\label{7}
b_0=11-{2\over3}n_f
\end{equation}
which gives $v=\alpha_s(m v)$ as a function of $m/\lqcd$. In Fig.~\ref{fig:3},
$\lqcd/ m \alpha_s^2(mv)$ is plotted as a function of $m/\lqcd$,
with $v=\alpha_s(m v)$. Clearly, for large $m/\lqcd$, one can have
$\lqcd \ll m v^2 \ll m v \ll m$.  However, it is not possible to have
$\lqcd \gg m v^2$. The maximum
possible value of $\lqcd/m v^2$ is
\begin{equation}\label{8}
{\lqcd \over m v^2} = {b_0 \over 2 \pi e}=0.53\ \ (\hbox{for }n_f=3), 
\end{equation}
at
\begin{equation}\label{9}
{m \over \lqcd}= {e b_0 \over 2 \pi}=3.9,\ v={2 \pi \over b_0}=
0.70\ \ (\hbox{for }n_f=3).
\end{equation}
[Here $e=2.718$.]  
(Of course these values should only be taken as illustrative, since if
$\lqcd\sim mv^2$ the system is no longer Coulombic). 
For the $J/\Psi$, $m_c/\lqcd \sim 3$ and for the
$\Upsilon$, $m_b/\lqcd \sim 9$ so that $m v^2/\lqcd$ is not very
different in the two cases.

\section{The Effective Theory}\label{sec:eft}

\begin{figure}
  \epsfxsize=6cm \hfil\epsfbox{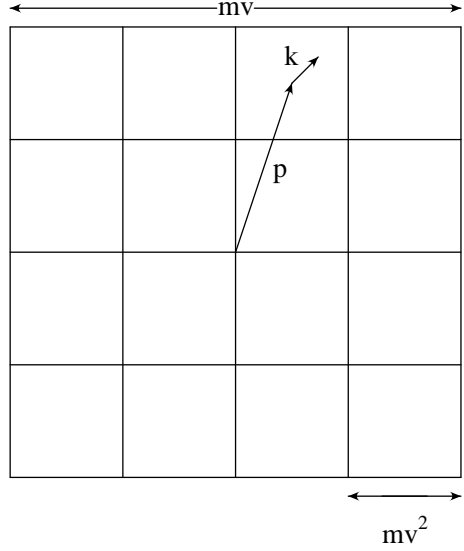}\hfill
\caption{Momentum space of size $mv$ is divided into boxes of size $mv^2$. A
  point in momentum space is labeled by $\mathbf p$ and $\mathbf k$.
\label{fig:4}}
\end{figure}

To construct the effective theory, label the total energy $E$ and
momentum $\mathbf{P}$ of the heavy quark by
\begin{equation}\label{eft:1}
{\mathbf P = p + k},\qquad E = k^0,
\end{equation}
where the three-vector $\mathbf{p}$ is of the order of the soft scale
$m v$, and the four-vector $k$ is of the order of the ultrasoft scale
$m v^2$.  Momentum space of size $m v$ is divided up into boxes of
size $m v^2$.  The location of each box is labeled by $\mathbf{p}$,
and the points within a box are labeled by $\mathbf{k}$, as shown in
Fig.~\ref{fig:4}
The variable $\mathbf{p}$ is a discrete label, and $\mathbf{k}$ is a
continuous label. This procedure was originally used by Georgi for
HQET~\cite{Georgi:1990}, where the four-momentum $p^\mu$ was split
between $m v^\mu$ of order $m$ and the residual momentum $k^\mu$ of
order $\lqcd$,
\begin{equation}\label{eft:3}
p^\mu=m v^\mu + k^\mu.
\end{equation}
In HQET, the velocity $v$ is a discrete label, and $k$ is a continuous
label, so that one sums on $v$ and integrates over
$k$~\cite{Georgi:1990}. In our case, we will sum over $\mathbf p$ and
integrate over $\mathbf k$.

The quark field $\psi(x)$ in QCD is replaced by
\begin{equation}\label{eft:2}
\psi\left(x\right) \to \psip p \left(x\right).
\end{equation}
The label $\mathbf{p}$ represents momenta of order the soft scale
$mv$, and (the Fourier transform of) $x$ represents energy and momenta
of order the ultrasoft scale $m v^2$. 
\OMIT{One might naively think that the
breakup Eq.~(\ref{eft:1}) represents a double counting of momentum,
but this is not the case. This should be clear from the similar
breakup~Eq.~(\ref{eft:3}), which is known not to cause any problems.}

The decomposition Eq.~(\ref{eft:3}) is not unique, since one can
redefine $k \to k + q$, $mv \to mv -q$, where $q$ is of order $k$.
This redefinition, called reparameterization invariance, leads to
constraints on the effective field theory, and relates different
orders in the $1/m$ expansion~\cite{Luke:1992a}. One can make a
similar redefinition here,
\begin{equation}\label{eft:5}
\mathbf{k} \to \mathbf{k+q},\qquad \mathbf{p} \to \mathbf{p-q},
\end{equation}
where $\mathbf{q}$ is of order $m v^2$. In terms of fields, this
transformation is
\begin{equation}\label{eft:4}
\psip p \left(x\right) \to e^{i \mathbf{q \cdot x} }\psip {p-q} \left(x \right).
\end{equation}
The application of reparameterization invariance to spinors in HQET
was subtle, because of the constraint $\xslash v \psi = \psi$, that
projected out the particle component of the spinor. In our case $\psip
p$ is a two-component spinor whose upper and lower components
represent the amplitudes to annihilate a quark with spin $\pm 1/2$
along a {\sl fixed} axis. The transformation Eq.~(\ref{eft:5}) does
not affect the spin labels, so the consequences of reparameterization
invariance are similar to the case of HQET for spin-zero particles.
[Spin would enter if the components of $\psip p$ represented helicity
states.] The basic result is that derivatives on $\psip p(x)$ should
be of the form $i\mathbf{p+\nabla}$~\cite{Luke:1992a}.

On-shell gauge fields have energy of order their momentum. One can
have propagating gauge fields with energy and momentum of order $mv$
or of order $mv^2$, which are referred to as soft and ultrasoft gauge
fields, respectively.  The gauge fields in the full theory are
replaced by two different fields in the effective theory,
momentum-dependent gauge fields, $A_{p}^\mu(x)$, and
momentum-independent gauge fields $A^\mu(x)$. The fields
$A_{p}^\mu(x)$ represent the soft degrees of freedom and $A^\mu(x)$
represent the ultrasoft degrees of freedom. The total energy and
momentum of the soft gauge fields is
\begin{equation}\label{eft:10}
{\mathbf P = p + k},\qquad E = p^0+k^0,
\end{equation}
and of the ultrasoft gauge fields is $k^\mu$, where $k$ is the Fourier
transform of the spacetime argument $x$. Note that soft gauge fields
are labeled by a four-vector $p$, whereas quark fields are labeled by
a three-vector $\mathbf p$. Any other light modes (such as light
fermions and ghosts) in the theory must also be divided into soft and
ultrasoft fields, as for the gauge fields.

The terms in the NRQCD effective
Lagrangian describe the interactions of the soft gauge fields among
themselves, and the interaction of two or more soft gauge fields with
the fermions. There are no terms that involve the interaction of a
fermion with a single soft gauge field, i.e.\ no vertex of the form
$\psi^\dagger_{\mathbf p^\prime} A^\mu_{q} \psi_{\mathbf p}$, since
energy cannot be conserved in the interaction.

The effective Lagrangian for NRQCD can now be written down in terms of the
fields $\psip p$ which annihilate a quark, $\chip p$ which annihilate an
antiquark, $A^\mu_{p}$ which annihilate and create soft gluons, and
$A^\mu$ which annihilate and create ultrasoft gluons. The covariant derivative
is $D^\mu = \partial^\mu + i g A^\mu=(D^0,-\mathbf{D})$, so that
$D^0=\partial^0+ig A^0$, ${\mathbf D}={\mathbf \nabla}-ig{\mathbf A}$, and
involves only the ultrasoft photon fields. The effective Lagrangian is
gauge invariant under ultrasoft gauge transformations, those in which the gauge
parameter varies on a distance scale $1/(m v^2)$. The full gauge invariance of
the original Lagrangian is recovered by combining gauge invariance in the
effective theory with reparameterization invariance.

The effective Lagrangian in the center of mass frame is
\begin{eqnarray}\label{nrqcd:1}
{\cal L} &=&  
 -{1\over 4}F^{\mu\nu}F_{\mu \nu} + \sum_{p} \abs{p^\mu A^\nu_p -
 p^\nu A^\mu_p}^2 + \sum_{\mathbf p}
 \psip p ^\dagger   \Biggl\{ i D^0 - {\left({\bf p}-i{\bf D}\right)^2 
\over 2 m} \Biggr\} 
 \psip p  \nonumber \\
&&  - 4 \pi \alpha_s
\sum_{q,q^\prime\mathbf p,p^\prime}\Biggl\{{1\over q^0} 
 \psip {p^\prime} ^\dagger  \left[A^0_{q^\prime},A^0_q \right]
  \psip p \nonumber \\
&&  + {g^{\nu 0} \left(q^\prime-p+p^\prime\right)^\mu -
  g^{\mu 0} \left(q-p+p^\prime\right)^\nu + g^{\mu\nu}\left(q-q^\prime \right)^0
  \over \mathbf \left( p^\prime-p \right)^2}
  \psip {p^\prime} ^\dagger  \left[A^\nu_{q^\prime},A^\mu_q \right]
  \psip p \Biggr\}\nonumber \\
 &&\qquad\qquad + \psi \leftrightarrow \chi,\ T \leftrightarrow \bar T \nonumber \\
&& + \sum_{\mathbf p,q}
  {4 \pi \alpha_s  \over \mathbf \left( p-q \right)^2} 
  \psip q ^\dagger T^A \psip p \chip {-q}^\dagger  \bar T^A \chip {-p} + \ldots
\end{eqnarray}
where we have retained the lowest order terms in each sector of the theory. The
matrices $T^A$ and $\bar T^A$ are the color matrices for the $\bf 3$ and $\bf
\bar 3$ representations, respectively. The field strength tensor $i g
F^{\mu\nu}=\left[D^\mu,D^\nu\right]$ is constructed only out of ultrasoft gauge
fields. If we were interested in including the effects of light quarks we would
need to add new soft and ultrasoft fields for each flavor. The matching at the
scale $m$ would then introduce additional  non-local four fermion terms
involving both the heavy and light quark fields.

In using Eq.~(\ref{nrqcd:1}), it is crucial to expand out the term
$\left({\bf p} - i {\bf D} \right)^2$. The ${\mathbf p}^2$ piece is part
of the leading order Lagrangian that gives
the $\psip p$ propagator,
\begin{equation}\label{nrqcd:4} 
\psip p ^\dagger   \Biggl\{ i D^0 - {{\bf p}^2
\over 2 m} \Biggr\} \psip p ,
\end{equation}
and the terms involving $\mathbf D$ are treated as a perturbation. 
This is the momentum space equivalent of the multipole expansion
written in $x$ space in \cite{Grinstein:1998}, since the ultrasoft
gluons do not transfer three momentum to the quarks.  
This
procedure will be justified when the velocity power counting rules for the
Lagrangian are derived. 

The QCD Lagrangian contains gauge fixing terms and ghost interactions. It is
convenient to quantize the theory in background field gauge. The background
fields can be taken to be the ultrasoft modes of the effective theory. The
quantum fields represent the quark potential, and the soft modes of the
effective theory. The effective Lagrangian is gauge invariant with respect
to the ultrasoft modes, and contains the gauge fixing terms of the original
theory for the soft modes. One can then gauge fix the ultrasoft gauge fields,
to compute loop graphs involving the ultrasoft gauge fields. We will use
Feynman gauge for both the soft and ultrasoft modes.

\begin{figure}
\epsfxsize=10cm
\hfil\epsfbox{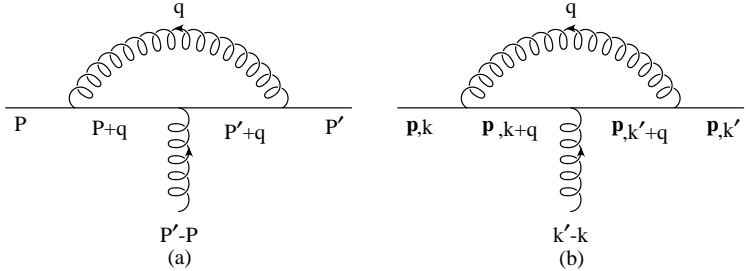}\hfill
\caption{One loop vertex correction in the full theory and effective theory. The
momenta are related by $P=p+k$, $P^\prime=p+k^\prime$.
\label{fig:X1}}
\end{figure}

The terms in the effective Lagrangian Eq.~(\ref{nrqcd:1}) are obtained by
matching graphs in QCD and the effective theory. The ultrasoft gluon fields
$A^\mu$ in the effective Lagrangian cannot change the momentum label $\mathbf
p$ on the fermion lines, so the single-quark terms in $L$ do not change
$\mathbf p$. The Lagrangian in the single-quark sector is the same as the HQET
Lagrangian, as pointed out in Ref.~\cite{Manohar:1997}. This relation between
the NRQCD and HQET Lagrangians holds even if loop corrections are included.
Consider a loop contribution to a term in the single-quark sector of the
effective theory, such as that shown in Fig.~\ref{fig:X1}.
In the effective theory, the incoming and outgoing momenta of the quark, $P$
and $P^\prime$ respectively, are broken up into a soft piece $\mathbf p$ and an
ultrasoft piece $k,k^\prime$. The soft label $\mathbf p$ must be the same on
the incoming and outgoing lines, since the ultrasoft gluons do not change
$\mathbf p$, so the momentum transfer in the effective theory is
$k^\prime-k=P^\prime-P$. The matching condition can be computed as the
difference between Fig.~\ref{fig:X1}(a) and (b), expanded in a power series in
$1/m$.  The full-theory computation is given by computing Fig.~\ref{fig:X1}(a)
on-shell, and expanding in powers of the external momenta over $m$. The
effective theory contribution is given by evaluating Fig.~\ref{fig:X1}(b)
on-shell. The on-shell condition in the effective theory is $k^0={\mathbf
p}^2/2m$. The intermediate fermion propagator
\begin{equation}\label{nrqcd:2}
{1\over k^0+q^0-{\mathbf p}^2/2m + i\epsilon}
\end{equation}
is equal to
\begin{equation}\label{nrqcd:3}
{1\over q^0 + i\epsilon}
\end{equation}
when the on-shell condition is used.\footnote{Note that we have used the lowest
order Lagrangian~Eq.~(\ref{nrqcd:4}) to derive the propagator, so the energy
term in the denominator involves the loop momentum $q$, but the momentum term
${\mathbf p}^2/2m$ does not.} The Feynman rules Eq.~(\ref{nrqcd:3}) are
precisely those that would be used to match from QCD to HQET, and are known not
to violate the $1/m$ power counting in the effective theory. Thus the couplings
of the ultrasoft gluons in the single-quark sector are precisely those in HQET.
This was the procedure used to compute the HQET and NRQCD Lagrangians at 
one-loop in Ref~\cite{Manohar:1997}.

\begin{figure}
\epsfxsize=6cm
\hfil\epsfbox{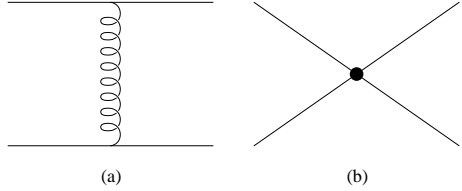}\hfill
\caption{The Coulomb potential in the full theory, (a), is given by a local 
two-quark operator (b) in the effective theory.
\label{fig:X2}}
\end{figure}

The Coulomb potential can scatter quark states from one value of $\mathbf p$ to
another. These effects are explicitly included in the two-body terms in $L$.
The Coulomb potential is usually thought of as a nonlocal two-body operator.
However, because of the use of the extra label $\mathbf p$, the Coulomb
potential,
\begin{equation}
{4 \pi \alpha \over \mathbf \left( p-q
 \right)^2} \psip q^\dagger T^A \left( x \right)
\psip p \left( x \right)
\chip {-q}^\dagger \left( x \right) \bar T^A 
\chip {-p} \left( x \right)
\end{equation}
is local, and manifestly gauge invariant. The Coulomb potential is obtained as
the value of Fig.~\ref{fig:X2} evaluated in the full theory.
The Coulomb potential is proportional to the Casimir $T^A\bar T^A$, and gives
an attractive interaction in the color singlet channel and a repulsive
interaction in the color octet channel.

\begin{figure}
\epsfxsize=12cm
\hfil\epsfbox{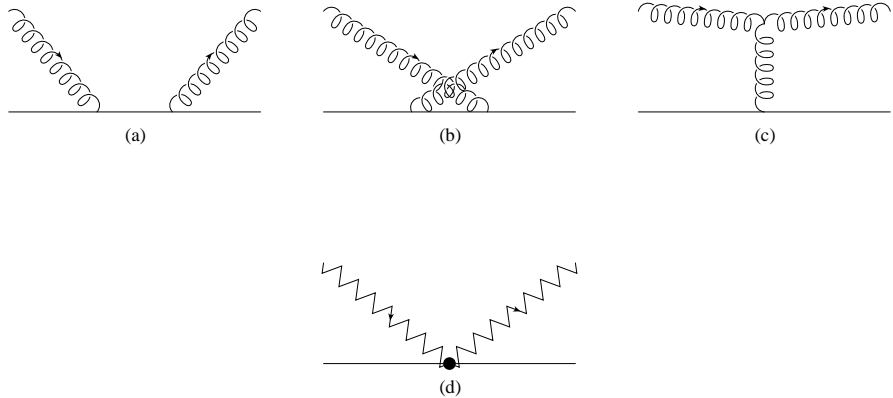}\hfill
\caption{The Compton scattering graphs (a,b,c) match on to a local operator (d)
in the effective theory. Soft gluon modes are denoted by a zigzag line.
\label{fig:X3}}
\end{figure}

The leading terms involving the soft gluons are given by matching the Compton
scattering graphs Figs.~\ref{fig:X3}(a,b,c) in the full theory to the local
operator  Fig.~\ref{fig:X3}(d) in the effective theory.
Soft gluons have energy and momenta of order $mv$, whereas the quarks $\psip p$
have energies of order $mv^2$ and momenta of order $mv$. The intermediate quark
in Fig.~\ref{fig:X3}(a,b) is off-shell, and the Compton scattering graph in the
full-theory can be replaced by a local vertex in the effective theory, as shown
in the figure. The intermediate gluon in Fig.~\ref{fig:X3}(c) is also
off-shell, since energy cannot be transferred from the external gluons to the
quark line. Thus the interaction in Fig.~\ref{fig:X3}(c) can also be
represented by a local vertex in the effective theory. The leading order soft
interaction vanishes for QED. We will comment on this in Sec.~\ref{sec:example}.
Loop graphs involving the soft interaction of Fig.~\ref{fig:X3}(d) are part of
the running potential in the effective theory.

The Lagrangian Eq.~(\ref{nrqcd:1}) is similar to the pNRQCD Lagrangian
constructed by Pineda and Soto in Ref.~\cite{Pineda:1998b}, but there are a few
important differences. The Lagrangian is given by matching to QCD at the scale
$\mu=m$, rather than at the scale $\mu=mv$. The Lagrangian also contains
explicit soft modes. Soft modes are necessary to reproduce the running
potential in the effective theory. Finally, the Coulomb potential is
constructed using a $\mathbf p$-dependent, but local in $\mathbf x$ two-body
operator, instead of a two-body operator nonlocal in $\mathbf x$. The use of
$\mathbf p$-dependent fields is the momentum space analog of the multipole
expansion, and simplifies the discussion of gauge invariance, particularly in
the non-Abelian case.

\subsection{Power Counting in the Lagrangian}\label{sec:pc}

The NRQCD effective Lagrangian can be used to compute processes to a given
order in the velocity $v$.  To determine the order in $v$ of a given diagram,
it is useful to have a velocity power counting scheme, and we will use the one
in Table~\ref{tab:2}.
\begin{table}
\caption{Velocity counting rules for the effective theory. The electric and
magnetic fields are those constructed out of the ultrasoft gauge potential
$A^\mu$.\label{tab:2}}
\begin{tabular}{ll}
${\mathbf p}$ & $v$ \\
$\psi,\ \chi$ & $v^{3/2}$ \\
$A_{p}^\mu$ & $v$ \\
$ D^0 $ & $v^2$ \\
$ \mathbf D $ & $v^2$ \\
$ A^\mu $ & $v^2$ \\
$ {\mathbf E} $ & $v^4$ \\
$ {\mathbf B} $ & $v^4$ \\
\end{tabular}
\end{table}
This velocity power counting scheme differs somewhat from that in BBL, since we
have separated the gluon field into soft and ultrasoft modes, and the Coulomb
interaction. The order in $v$ of the soft gluons modes $A_p$ is irrelevant,
since they cannot appear as external states in the processes we are
considering. The final power counting formula for graphs we will derive in
Eq.~(\ref{nrqcd:21}) holds regardless of the order in $v$ chosen for $A_p$. The
lowest dimension operator in the zero-quark sector of the Lagrange density is
the gauge kinetic energy, which is of order $v^8$. All the terms in the field
strength tensor $F^{\mu\nu}$ are of the same order in $v$, since $D^\mu$ and
$A^\mu$ are both of order $v^2$.

The lowest dimension terms in the one-quark sector are
\begin{equation}\label{pc:1}
\psip p ^\dagger \left( x \right) \left\{ i D^0 - {{\bf p}^2
\over 2 m} \right\}\psip p  \left( x \right)
\end{equation}
which are of order $v^5$. The lowest order Lagrangian Eq.~(\ref{pc:1}) must be
used to determine the $\psip p$ propagator. Terms that involve the covariant
derivative $\mathbf D$ are of higher order than $v^5$. For example,
\begin{equation}\label{pc:2}
\psip p ^\dagger {{\mathbf p \cdot D}
\over  m}\psip p
\end{equation}
is of order $v^6$, and must be treated as a perturbation in the effective
theory. Each replacement of $\mathbf p $ by $\mathbf D$ increases the order in
$v$ by one.

The lowest dimension operator in the two-quark sector is the Coulomb
interaction. Each quark field is of order $v^{3/2}$
and $1/({\mathbf p - q})^2$ is of order $1/v^2$, so the Coulomb interaction is
of order $\alpha_s v^4$.

The lowest dimension terms that
involve soft gluons are of order $\alpha_s v^4$. Consider, for example,
\begin{equation}
 - 4 \pi \alpha_s
\sum_{q,q^\prime\mathbf p,p^\prime}{1\over q^0} 
 \psip {p^\prime} ^\dagger  \left[A^0_{q^\prime},A^0_q \right]
  \psip p .
\end{equation}
The two $\psip p$ fields are of order $v^3$, the two $A^0_q$ fields 
are of order $v^2$, and
$1/q^0$ is of order $1/v$, so the vertex is of order $\alpha_s v^4$.

\subsection{Loop Graphs}\label{sec:loop}

The computation of loop graphs using the effective
Lagrangian~Eq.~(\ref{nrqcd:1}) involves some subtleties. There are three kinds
of loops, which we will refer to as ultrasoft, potential and soft,
respectively. We will determine the dominant momentum region for each graph by
studying the pole structure of the diagram, which is what determines the
behavior of the graph in the \ms\ scheme.

Consider graphs in the effective theory that involve only a single fermion
line, such as Fig.~\ref{fig:X1}(b). The internal fermion propagators are given by
the lowest order term in the one-fermion sector, Eq.~(\ref{nrqcd:2}). These
terms depend only on the soft momentum $\mathbf p$, and not on the ultrasoft
momentum $\mathbf k$ carried by the gluon lines. Thus the two fermion
propagators in Fig.~\ref{fig:X1}(b) are
\begin{equation}
{1\over k^0 +q^0- {\bf p}^2/2m},\ \hbox{and}\ 
{1\over k^{\prime 0} + q^0- {\bf p}^2/2m},
\end{equation}
respectively. These propagators do not depend on the space part of the loop
momentum $\mathbf q$, but they do depend on $m$. The propagators have poles at
energies of order ${\bf p}^2/2m$ or $k^0$. The pole positions are set
by the external gluon and quark energies, and also by ${\bf p}^2/2m$, which
depends on the quark momentum. The fact that the energy poles are determined by
the momentum (and vice versa), is important, because it relates the soft and
ultrasoft scales. We will refer to a typical external quark energy as of the 
order of the ultrasoft scale $\muus$, and a typical external quark momentum as
of the order of the soft scale $\mus$. Then loop graphs in the one-fermion
sector have poles at energies of order $\muus$ and $\mus^2/m$. The gauge boson
propagator is $1/[(q^0)^2-{\mathbf q}^2]$, so the typical momenta in the loop
are of the same order as the gluon energy, $\mathbf q \sim q^0$. Loop graphs
such as Fig.~\ref{fig:X1}(b) in which the ultrasoft momentum $q$ is integrated
over will be referred to as ultrasoft loops, and when evaluated in dimensional
regularization, are dominated by energy and momenta of order $\muus$ or
$\mus^2/m$.

The use of ${\bf p}^2/2m$ rather than $\left({\bf p+q}\right)^2/(2m)$ in the
propagators removes the problem of the breakdown of the effective theory due to
poles of order $m$ in loop graphs. The power counting scheme of
Table~\ref{tab:2} in which $\mathbf p$ is of order $m v$, but $\mathbf q$ is of
order $m v^2$ requires that $\left({\mathbf p+q}\right)^2$ be expanded as
$\mathbf p^2 + 2 p \cdot q + q^2$, with $\bf p^2$ included in the fermion
propagator, and the higher order terms  $\mathbf 2 p \cdot q + q^2$ treated as
vertex insertions. 

\begin{figure}
\epsfxsize=10cm
\hfil\epsfbox{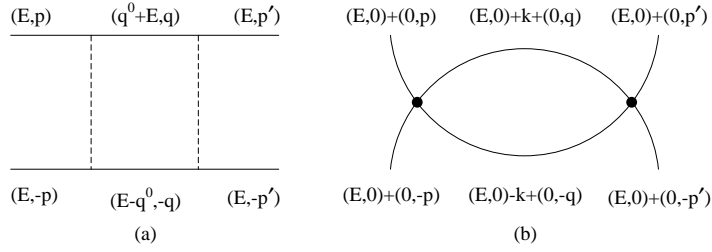}\hfill
\caption{One loop correction to Coulomb Scattering in the full and effective
theories.
\label{fig:6}}
\end{figure}

Consider a graph that involves a potential loop, such as the one-loop
correction to Coulomb scattering. The graph is shown as Fig.~\ref{fig:6}(a)
in the full theory, and as Fig.~\ref{fig:6}(b) in the effective theory.
The external fermions have energy $E$ and momenta $\pm \mathbf p,p^\prime$. In
the effective theory, the external fermions are labeled by the soft momentum 
$\pm \mathbf p,p^\prime$, and the ultrasoft momentum $(E,0)$. The intermediate
fermions in the effective theory have soft momentum $\pm q$, and ultrasoft
momentum $(E,0)\pm k$. The graph in the effective theory is
\begin{equation}\label{nrqed:5}
\propto \sum_{\mathbf q} \int d^4 k {1 \over \mathbf (p-q)^2}\
{1 \over \mathbf (p^\prime-q)^2}\ {1 \over k^0+E - {\mathbf q}^2/2m}\
{1 \over -k^0+E - {\mathbf q}^2/2m}.
\end{equation}
There is an integral over the ultrasoft energy $k^0$, the ultrasoft momentum
$\mathbf k$, and a sum over the soft momentum $\mathbf q$. Recall the
decomposition of momentum space shown in Fig.~\ref{fig:4}. Summing over 
$\mathbf q$ and integrating over $\mathbf k$ is equivalent to integrating
over the entire momentum space. Thus one can replace Eq.~(\ref{nrqed:5}) by
\begin{equation}\label{nrqed:6}
\int d^4 q {1\over \mathbf (p-q)^2}
{1\over \mathbf (p^\prime-q)^2} {1 \over q^0+E - {\mathbf q}^2/2m}
{1 \over -q^0+E - {\mathbf q}^2/2m},
\end{equation}
where 
\begin{equation}\label{nrqed:11}
dk^0 \to d q^0,\qquad \sum_{\mathbf q} \int d{\mathbf k} \to 
\int d{\mathbf q}.
\end{equation}
The loop graph is dominated by $q^0\sim E$, and  ${\mathbf q} \sim \sqrt{m E}$,
which are ultrasoft, and soft, respectively. This is consistent with the
picture that $q^0$ ranges over a box of size $m v^2$ and $\mathbf q$ over a box
of size $m v$. It again shows that one cannot treat the soft and ultrasoft
scales as independent of each other; they are related by $\hbox{(soft)}^2 \sim
\hbox{ultrasoft}\times m$.

\begin{figure}
\epsfxsize=4cm
\hfil\epsfbox{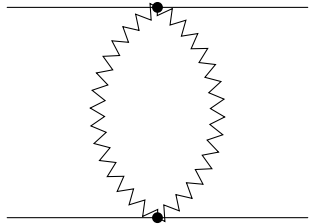}\hfill
\caption{One loop graph involving the soft gauge fields. This graph contributes
to the running potential in the effective theory.
\label{fig:X5}}
\end{figure}

Finally, consider a graph such as Fig.~\ref{fig:X5} that involves a soft loop.
It gives a contribution of the form
\begin{equation}\label{nrqcd:8}
\propto \sum_q \int d^4 k {1\over \mathbf (p-p^\prime)^2}\
{1 \over \mathbf (p-p^\prime)^2}\ {1 \over \left(q^0\right)^2-{\mathbf q}^2}\
{1 \over \left(q^0\right)^2-\left({\mathbf q+p-p^\prime}\right)^2},
\end{equation}
where we have used the last interaction in Eq.~(\ref{nrqcd:1}) for the soft
vertices. The sum on $q$ is over a four-vector. As for
potential loops, one can make the replacement 
\begin{equation}\label{nrqed:11s}
\sum_{q} \int d^4{ k} \to 
\int d^4{q},
\end{equation}
where the replacement must be done for all four components of $q$, since soft
gluons carry a four-vector label $q$. It is straightforward to see that
Eq.~(\ref{nrqcd:8}) is dominated by $q^0$ and $q$ of order $mv$, which is
consistent with using the replacement Eq.~(\ref{nrqed:11}) for all four
components of $q$.

\subsection{Power Counting Formula for Loop Graphs}

\begin{figure}
\epsfxsize=6cm
\hfil\epsfbox{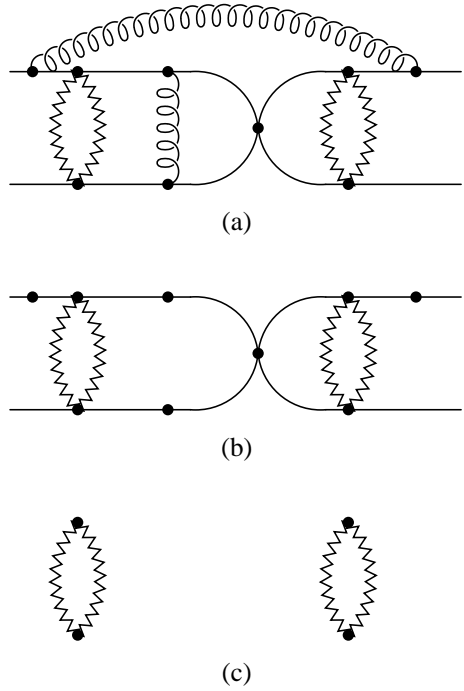}\hfill
\caption{An example of counting ultrasoft, potential and soft loops and
vertices. Graph (a) has 9 vertices and 6 loops. Deleting all ultrasoft lines
gives graph (b), which has 9 vertices and 4 loops. Deleting all fermion lines
from graph (b) gives graph (c), with 4 vertices and 2 loops, which is the
number of soft vertices and loops. There are 5 potential vertices and 2
potential loops (the difference between (b) and (c)), and 4 ultrasoft vertices
and 2 ultrasoft loops (the difference between (a) and (b)). The graph (c) has
two connected soft components, $N_S=2$.
\label{fig:X6}}
\end{figure}

It is now straightforward to derive a power counting rule for an arbitrary
graph in NRQCD. A given graph has $L_{U}$ ultrasoft loops,
$L_P$ potential loops, and $L_S$ soft loops. These can be determined from the
structure of the diagram in a systematic way. The total number of loops is
$L_U+L_P+L_S$. Now delete all the ultrasoft lines from the graph. The remaining
number of loops is $L_P+L_S$. Finally, delete all quark lines from the diagram.
The remaining number of loops is $L_S$. An example of loop counting is shown in
Fig.~\ref{fig:X6}.

Let $V_k$ denote the number of vertices of order $v^k$ in a given graph.
For example, $\psip p^\dagger iD^0 \psip p$ is a vertex of type $V_5$, since it
is a one-fermion vertex of order $v^5$. It is convenient to break up the
vertices into ultrasoft, $V^{(U)}_k$, potential $V^{(P)}_k$ and soft 
$V^{(S)}_k$. The ultrasoft vertices $V^{(U)}_k$ are those vertices in $V_k$
that involve only ultrasoft fields, the potential vertices $V^{(P)}_k$ are
those with at least one fermion field and no soft fields. The soft
vertices $V^{(S)}_k$ involve at least one soft field. An example of vertex
counting is shown in Fig.~\ref{fig:X6}.

A diagram is of order $v^\delta$, where
\begin{equation}\label{nrqed:7}
\delta = \sum_{k} k \left[V^{(U)}_k + V^{(P)}_k  + V^{(S)}_k  \right]
- 5 I_F - 4 I_S - 8 I_U + 8 L_U + 5 L_P + 4 L_S.
\end{equation}
The first term simply adds up the powers in $v$ of all the vertices. Each
internal quark line eliminates two $\psip p$ fields, and gives a factor of the
fermion propagator $1/(E - {\bf p}^2/2m)$, which give a net factor of  $1/(v^3
v^2)=v^{-5}$. This gives the term $-5 I_F$, where $I_F$ is the number of
internal fermion lines. Each internal soft line eliminates two $A_p$ fields,
and gives a factor of the gauge propagator $1/((p^0)^2 - {\bf p}^2)$, which
gives a net factor of  $1/(v^2 v^2)=v^{-4}$. This gives the term $-4 I_S$,
where $I_S$ is the number of internal soft lines. Each internal  ultrasoft line
eliminates two $A^\mu$ fields, and gives a factor of the gauge propagator
$1/((k^0)^2 - {\bf k}^2)$, which gives a net factor of  $1/(v^4 v^4)=v^{-8}$.
This gives the term $-8 I_U$, where $I_U$ is the number of internal ultrasoft
lines. 

Ultrasoft loops are dominated by energy and momentum of order $mv^2$, and give
give a factor of $v^2$ for each integral, so that one gets a factor of $v^8$
for each loop. Potential loops are dominated by energy of order $mv^2$ and
momentum of order $mv$, and give a factor of $v^2$ for each time integration,
and $v$ for each space integration, for a net factor of $v^5$ per loop. Soft
loops are dominated by energy and momentum of order $mv$, and give a factor of
$v$ for each integration, for a net factor of $v^4$ for each loop. These
contributions give the $8 L_U + 5 L_P + 4 L_S$ term in Eq.~(\ref{nrqed:7}).

The identity
\begin{equation}\label{nrqed:9}
\sum_{k,n} \left[V^{(U)}_k + V^{(P)}_k  + V^{(S)}_k  \right] - I_F - I_S
-I_U + L_U + L_P + L_S =1,
\end{equation}
is the usual relation that the Euler character of a
connected graph is unity. An analogous relation holds for the graph with all
ultrasoft lines removed. We will assume that the graph remains connected
when ultrasoft lines are removed, which is true for any process in which 
momentum of order $mv$ is transferred between the two fermion lines. 
The relation for the graph with ultrasoft lines removed is
\begin{equation}\label{nrqed:10}
\sum_k \left[V^{(P)}_k  + V^{(S)}_k  \right]- I_F -I_S + L_P + L_S =1.
\end{equation}
For $I_F$ to be equal in the two relations Eq.~(\ref{nrqed:9}) and
(\ref{nrqed:10}), it is
important that one not erase vertices where the gluons couple to the fermions
(see Fig.~\ref{fig:X6}), so the total number of vertices is given by 
$V^{(P)}_k + V^{(S)}_k$. Finally, one has the Euler character relation for the
graph with all ultrasoft and fermion lines removed,
\begin{equation}\label{nrqed:10a}
\sum_k V^{(S)}_k -I_S + L_S =N_S,
\end{equation}
where $N_S$ is the number of connected components in the soft graph. 

Eliminating $I_U$, $I_F$ and $L_S$ between Eqs.~(\ref{nrqed:7}--\ref{nrqed:10a})
gives the result
\begin{equation}\label{nrqed:8}
\delta = 5 + \sum_k \left[(k-8) V^{(U)}_k+(k-5) V^{(P)}_k +(k-4) V^{(S)}_k\right] 
 -N_S .
\end{equation}
A given soft vertex in the Lagrange density has the generic form
\begin{equation}\label{nrqcd:20}
\left(\psip p\right)^a \left( A_q \right)^b \left( A^\mu \right)^c
{\mathbf p}^d \left(D \right)^e \left({1\over m} \right)^f,
\end{equation}
where $\psip p$ represents any quark or antiquark fields or their conjugates.
The term Eq.~(\ref{nrqcd:20}) has dimension four,
\begin{equation}\label{nrqcd:21c}
4={3\over2}a+b+c+d+e-f,
\end{equation}
and is of order $v^k$, where
\begin{equation}\label{nrqcd:21a}
k={3\over2}a+ b+2 c+d+2 e.
\end{equation}
Subtracting these two relations gives
\begin{equation}\label{nrqcd:21b}
k-4=c+e+f.
\end{equation}
The vertex can only have positive powers of $D$, $A^\mu$, and $1/m$, so
$c+e+f\ge0$. [Note that $d$ need not be positive.] For a soft vertex, let
$\sigma = c+e+f\ge 0$. Label the soft vertices $V^{(S)}_k$ by the values of $b$
and $\sigma$, so that they are denoted by $V^{(S)}_{b,\sigma}$, where
$k=4+\sigma$. Then one finds
\begin{equation}
\sum_k (k-4) V^{(S)}_k = \sum_{b,\sigma} \sigma  
V^{(S)}_{b,\sigma}.
\end{equation}
Substituting this result into Eq.~(\ref{nrqed:8}) gives the power counting
formula
\begin{equation}\label{nrqcd:21}
\delta = 5 + \sum_k \left[(k-8) V^{(U)}_k+(k-5) V^{(P)}_k \right]
+\sum_\sigma \sigma V^{(S)}_{b,\sigma} 
-N_S .
\end{equation}
This is an important result. All terms in the zero-fermion sector have $k\ge
8$, and in the nonzero fermion sector have $\sigma \ge 0$. Thus the
contributions of $(k-8) V^{(U)}_k$, and $\sigma V^{(S)}_{b,\sigma}$ are each
positive. There can be negative powers of $v$ from both soft and potential
exchange, but these come with compensating factors of $\alpha_s$. The Coulomb
interaction is $k=4$ vertex in $V^{(P)}$, but is of order $\alpha_s$. Iterating
the Coulomb interaction $n$ times produces terms of order $(\alpha_s/v)^n$, so
for $v\sim\alpha_s$, the Coulomb interaction must be summed to all orders. The
$N_S$ term is negative, but each soft component must contain at least one power
of $\alpha_s$.

\begin{figure}
\epsfxsize=4cm
\hfil\epsfbox{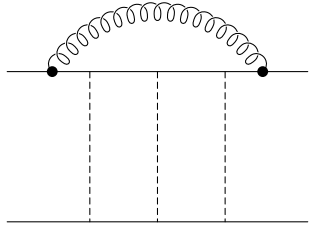}\hfill
\caption{Graph contributing to the Lamb shift. The dot represents a $\mathbf p
\cdot A$ interaction.
\label{fig:X7}}
\end{figure}

As an example of Eq.~(\ref{nrqed:8}), consider the electron self-energy diagram
Fig.~\ref{fig:X7} 
which produces the Lamb shift. The graph contains two $\mathbf p \cdot A$
vertices, each of which is of order $v^6$, and one ultrasoft loop, so the net
power is $\delta=5+(6-5)\times 2 =7$. The graph has a factor of $\alpha_s$ 
from the two gauge couplings, and so is of order $\alpha_s v^7$ compared to the
leading term in the single-fermion sector (which is of order $v^5$), i.e. it is
of relative order $\alpha_s v^2$. Similarly the soft loop graph in
Fig.~\ref{fig:X5} has $\sigma=0$, and $N_S=1$, and so is of order $\alpha_s^2
v^4$. This is the same order in $v$, but one higher order in $\alpha_s$ than
the Coulomb interaction.

\section{Velocity Renormalization Group Equation}\label{sec:VRG}

Loop diagrams in the effective theory can be divergent, and the effective theory
is renormalized using the \ms\ scheme. A scale parameter $\muus$ is introduced
when one analytically continues the Lagrangian from four to $D=4-2\epsilon$
dimensions, as in conventional dimensional regularization of gauge theories.
There is a subtlety in the use of dimensional regularization in NRQCD.
In evaluating
loop graphs with potential loops, we made the replacement Eq.~(\ref{nrqed:11}).
In $D$ dimensions, the relation should read
\begin{equation}\label{rg:1}
dk^0 \to d q^0,\qquad \sum_{\mathbf q}  \int d^{D-1}{\mathbf k} \to  
\left( {\mus\over \muus} \right) ^{4-D} \int d^{D-1}{\mathbf q}.
\end{equation}
The factor of $\left( {\mus/ \muus} \right) ^{4-D} $ is needed to ensure the
correct dimensionality on the two sides of Eq.~(\ref{rg:1}). The integral over
$d{\mathbf k}$ is over a volume of the order of $\muus^{D-1}$, since the 
typical range of integration
is of order $\muus$. The integral over $d{\mathbf q}$ is over a volume of 
the order of $\mus^{D-1}$,
since the range of integration is of order $\mus$. The number of terms in the
sum on ${\mathbf q}$ on the left-hand side of Eq.~(\ref{rg:1}) is the ratio of
the two volumes in four dimensions, $\left( {\mus/ \muus} \right) ^3 $. Away
from four dimensions, this number does not properly account for the momentum
space volumes on the two sides of Eq.~(\ref{rg:1}), and the additional factor of
$\left( {\mus/ \muus} \right) ^{D-4} $ is needed. The mismatch of dimensions in
$D\not=4$ occurs only for the space part of the integral. The factor above is
correct when the NRQCD integrals are done in the conventional way, by first
doing the $k^0$ integral using the method of residues disregarding the contour
at infinity, followed by the $\mathbf k$ integral in $3-2\epsilon$ dimensions.
Similarly, for soft loops Eq.~(\ref{nrqed:11s}) should be replaced by
\begin{equation}\label{rg:1s}
\sum_{q}  \int d^D{k} \to  
\left( {\mus\over \muus} \right) ^{4-D} \int d^D{q}.
\end{equation}

The effective theory renormalized in the \ms\ scheme has two $\mu$ parameters,
$\mus$ and $\muus$. However, the two parameters are not independent,  since the
soft and ultrasoft scales are related, $\mus^2 = m \muus$. It is better to
think of the parameters as $\mus=m \nu$ and $\muus=m \nu^2$, where $\nu$ is the
subtraction point velocity. One can now derive a new kind of renormalization
group equation for the effective theory, since the bare theory is independent
of the subtraction velocity $\nu$. This velocity renormalization group equation
can be used to scale the coefficients in the effective theory from the matching
scale $\mus=m$, $\muus=m$ to $\mus = mv$, $\muus=mv^2$, i.e.\ from $\nu=1$ to
$\nu=v$. 

The velocity renormalization group equation addresses an important point
about the effective theory, the simultaneous existence of two related scales
$\mus$ and $\muus$. Loop graphs involving gluon loops will typically have
logarithms of the form $\ln \mu/E$, where $E\sim mv^2$ is the typical photon
energy. The scale $\mu$ is equal to $\muus$, since $\muus$ is the scale
parameter introduced in $D$ dimensions. Loop graphs containing four-fermion
terms, (the potential and soft loop graphs discussed earlier), typically have
logarithms  of the form $\ln \mu/\sqrt{mE}$ or $\ln \mu/{\mathbf p}$. These
graphs use the replacement Eq.~(\ref{rg:1}), and so have have a factor of
\begin{equation}
\left( {\mus \over \muus} \right)^{2\epsilon} \muus^{2\epsilon} = 
\mus^{2\epsilon}
\end{equation}
for each loop, where the first factor is from Eq.~(\ref{rg:1}), and the second
factor is the conventional factor for each loop in dimensional regularization.
Thus in potential and soft  loops, logarithms are of the form $\ln
\mus/\sqrt{mE}$ or $\ln \mus/{\mathbf p}$.  The radiation and potential
logarithms are
\begin{equation}
\ln {\muus \over E} = \ln {m \nu^2 \over m v^2},\qquad
\ln {\mus \over \sqrt{mE}} = \ln {m \nu \over m v}.
\end{equation}
The choice of renormalization point $\nu=v$ ensures that both logarithms are
simultaneously small. Thus using the velocity renormalization group equation
from $\nu=1$ to $\nu=v$ simultaneously sums the logarithms involving the soft
and ultrasoft scales. The velocity renormalization group equation is the \ms\
equivalent of using an energy cutoff $m \nu^2$ and a momentum cutoff $m \nu$ in
a hard cutoff scheme. The VRG allows one to have a static potential with an
effective coupling constant $\alpha_s(m\nu)$, and radiation corrections with an
effective coupling constant $\alpha_s(m\nu^2)$.

In a conventional renormalization group approach, one would scale the effective
theory from $\mu=m$ to $\mu=mv$, and then down to $\mu=mv^2$. The VRG differs
from this in an important way, because it uses a subtraction velocity  rather
than a subtraction scale. Scaling the theory from $\nu=1$ to $\nu=v$ is
equivalent to simultaneously scaling potential and soft graphs from $m$ to
$mv$, and radiation graphs from $m$ to $mv^2$. The scale $mv$ and $mv^2$ are
coupled in the theory, and this coupling of scales is better treated using a
subtraction velocity rather than a subtraction scale.

\section{Examples}\label{sec:example}

The formalism we have developed will be applied to three illustrative examples 
in this section, the one-loop correction to the static potential, integrating
out a heavy fermion, and the two-loop anomalous dimension of the production
current~\cite{Hoang:1997a,Hoang:1997b}.

\subsection{Box Graph and the Static Potential}

The first example we consider is the renormalization of the static potential at
one-loop. At tree-level, the fermion-fermion scattering amplitude is reproduced
in the effective theory by a local operator, the Coulomb vertex in
Eq.~(\ref{nrqcd:1}). At one-loop, the QCD diagrams that contribute are shown in
Table~\ref{tab:box} above the horizontal line, and the graphs in the effective
theory are shown below the horizontal line.
\begin{table}
\caption{One-loop correction to quark-antiquark scattering.  The color factors
listed in the table are for the color singlet channel. The column labeled QCD
gives the contribution of the graph evaluated in QCD in the \ms\ scheme in
Feynman gauge, from Titard and Yndurain~\cite{Titard:1994,Titard:1995}. Only the
logarithmic contributions are given. $\lambda$ is a gluon mass used as an
infrared regulator, $k$ is the momentum transfer, and $\mu$ is the scale
parameter of dimensional regularization.  The gluon vacuum polarization graph
includes the contribution from the ghost loop. The column labeled HQET gives
the contributions of the diagrams in HQET, where $m \to \infty$. The breakup of
the total diagram into soft and ultrasoft contributions is given in the last
two columns. The two diagrams below the horizontal line are the
contributions to quark-antiquark scattering in the effective theory. The
complete set of ultrasoft diagrams is show in Fig.~\ref{fig:usbox}.
\label{tab:box}}
\[
\begin{array}{lc|cc|cc}
\hbox{Diagram} & \hbox{Color Factor} & \hbox{QCD} & \hbox{HQET} & \hbox{Soft} 
& \hbox{Ultrasoft}\\
\hline \hline &&&&&\\
\epsfxsize=2cm \lower15pt\hbox{\epsfbox{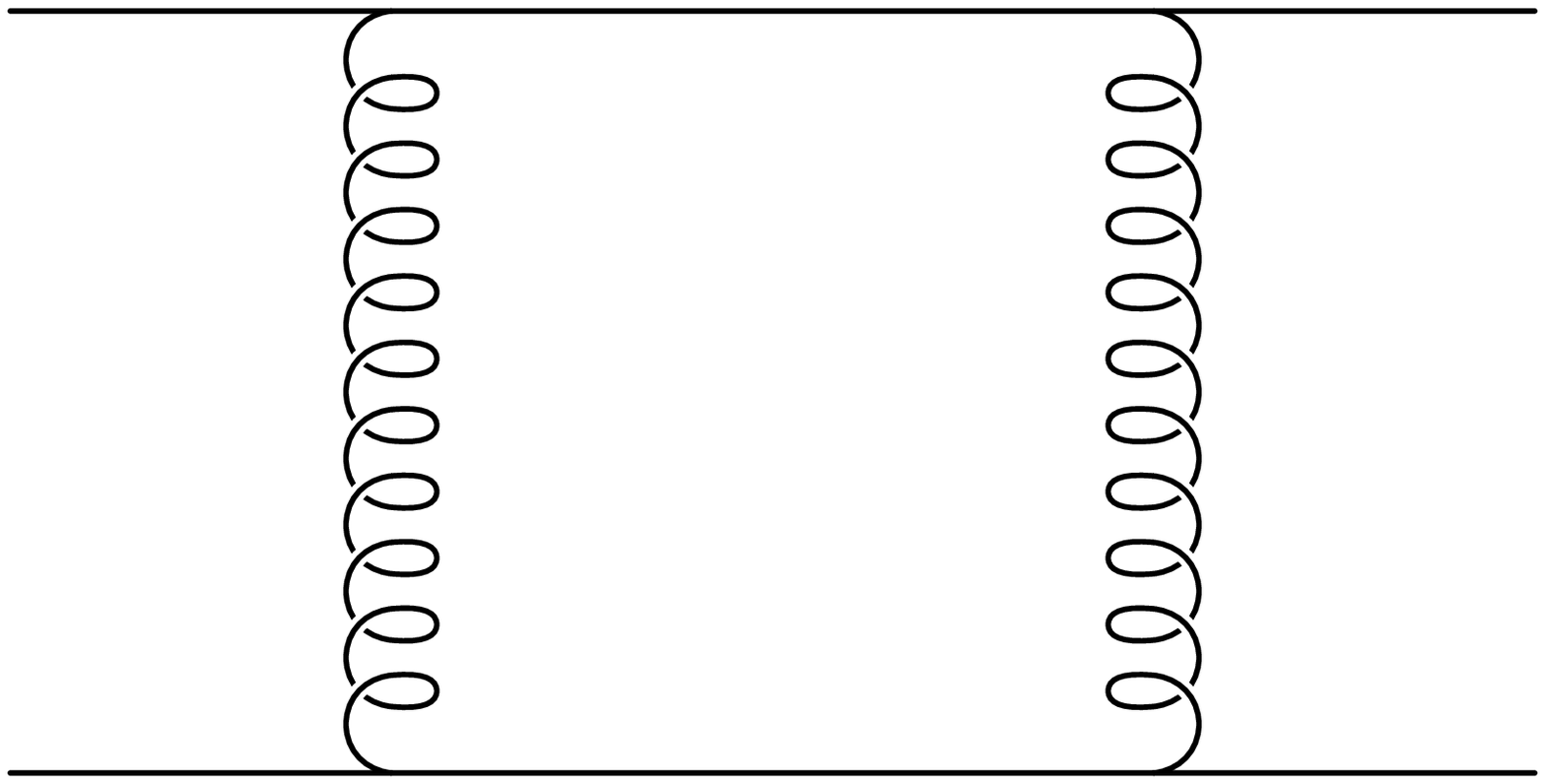}} & C_F &
2 \ln {\lambda \over k} &
2 \ln {\lambda \over k} & -2 \ln {k \over \mu } & 2 \ln {\lambda \over \mu}\\
&&&&&\\
\epsfxsize=2cm \lower15pt\hbox{\epsfbox{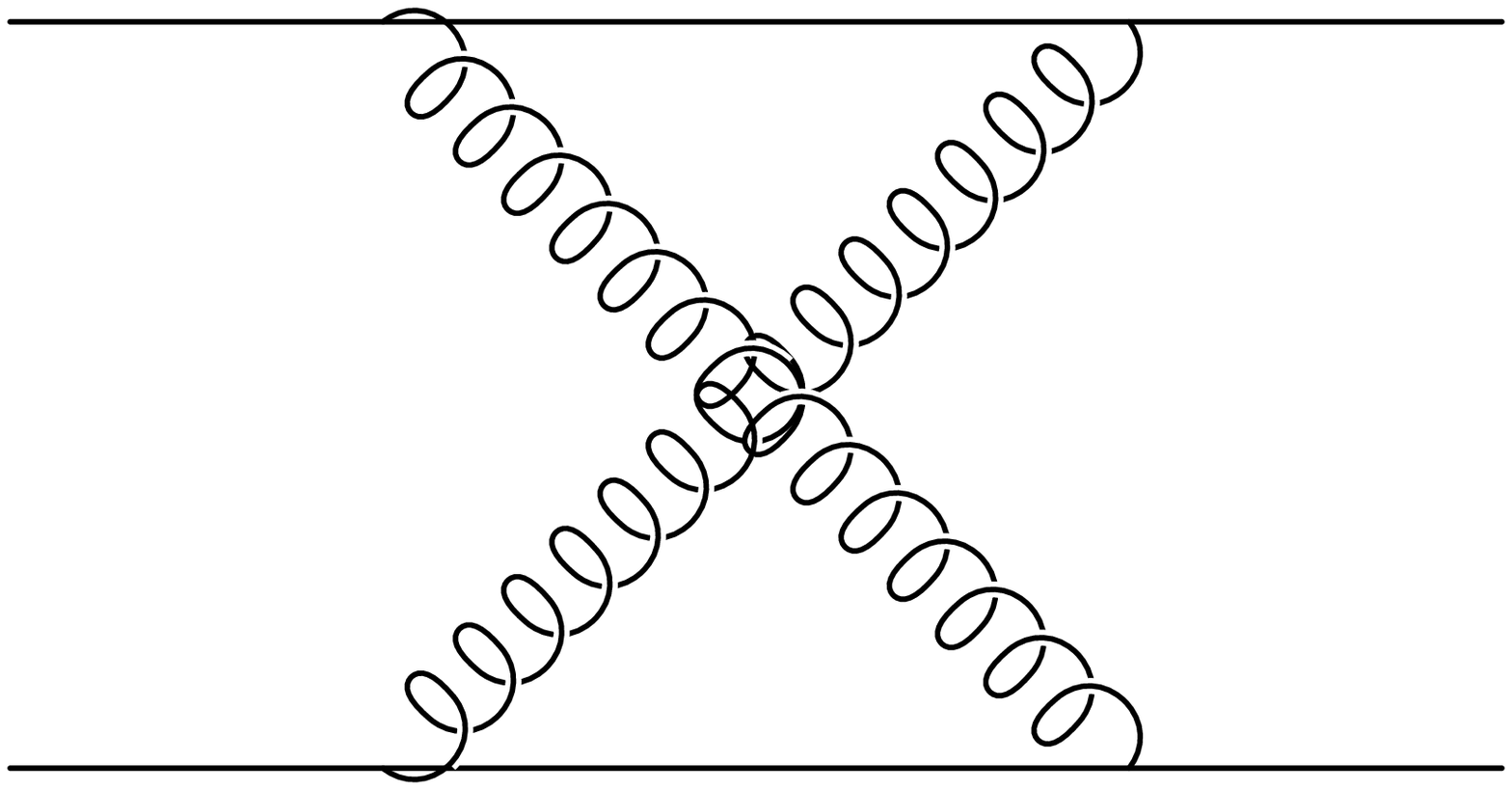}}  &
\left(C_F-{1\over2}C_A \right)  & -2 
\ln {\lambda \over k} & -2 
\ln {\lambda \over k} & 2 \ln {k \over \mu} & -2 \ln {\lambda \over \mu}\\
&&&&&\\
\epsfxsize=2cm \lower15pt\hbox{\epsfbox{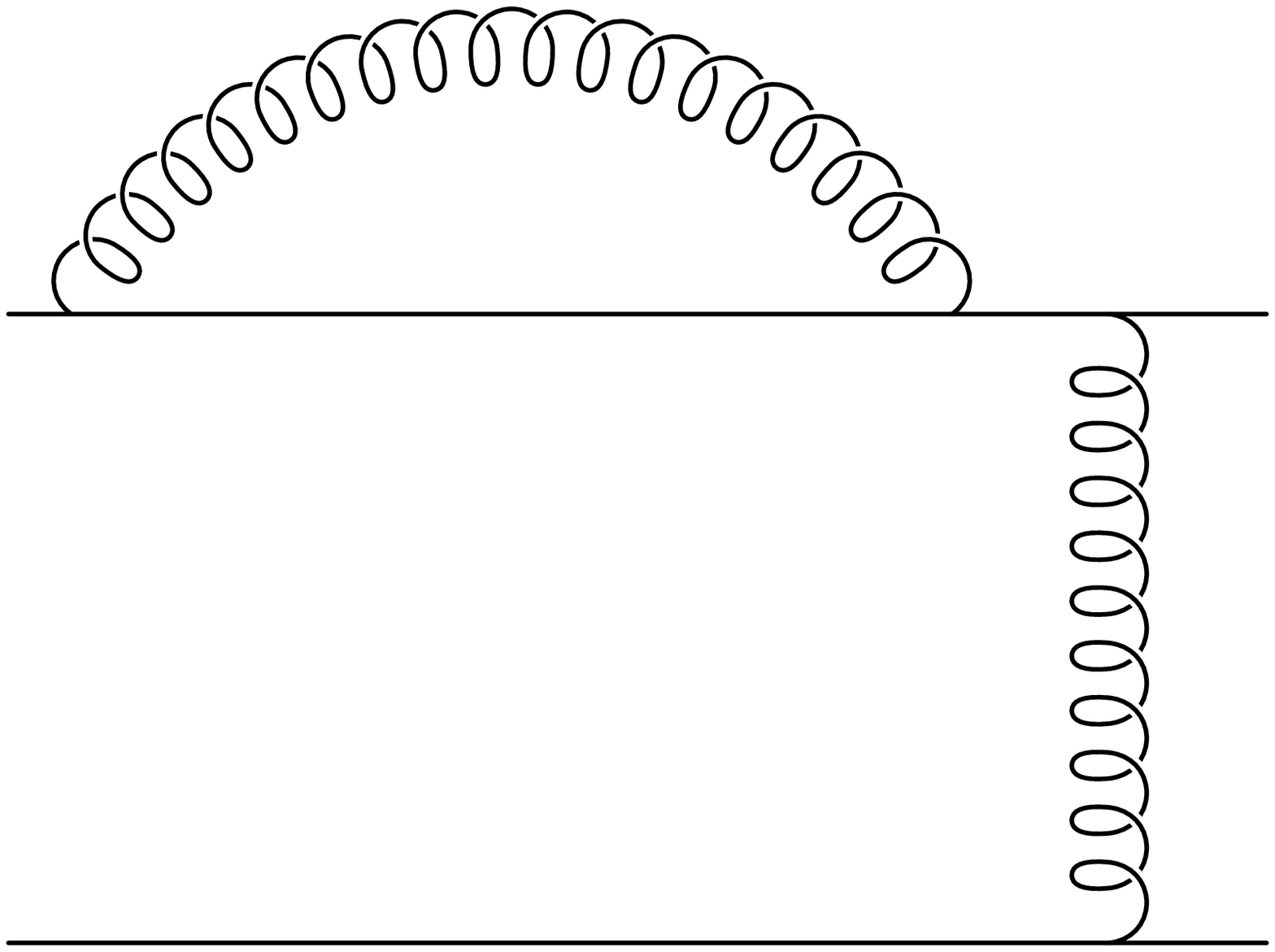}}  & C_F & 
- \ln {\mu \lambda^2 \over m^3} & 
- 2\ln {\lambda \over \mu} & 0 & - 2\ln {\lambda \over \mu}\\
&&&&&\\
\epsfxsize=2cm \lower15pt\hbox{\epsfbox{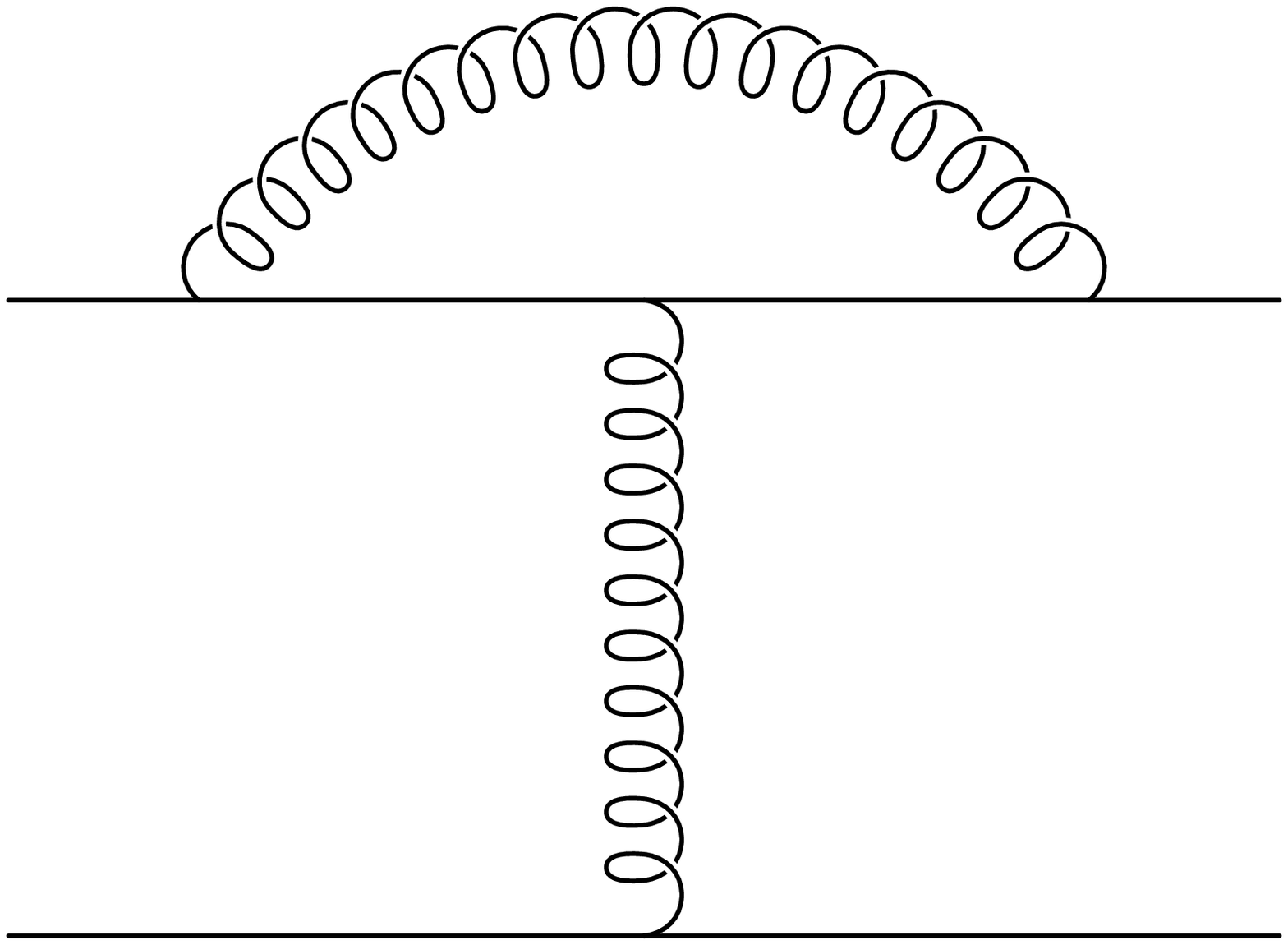}} &
\left(C_F-{1\over2}C_A \right) &
\ln {\mu \lambda^2 \over m^3} & 
2\ln {\lambda \over \mu} & 0 & 2 \ln {\lambda \over \mu}\\
&&&&&\\
\epsfxsize=2cm \lower15pt\hbox{\epsfbox{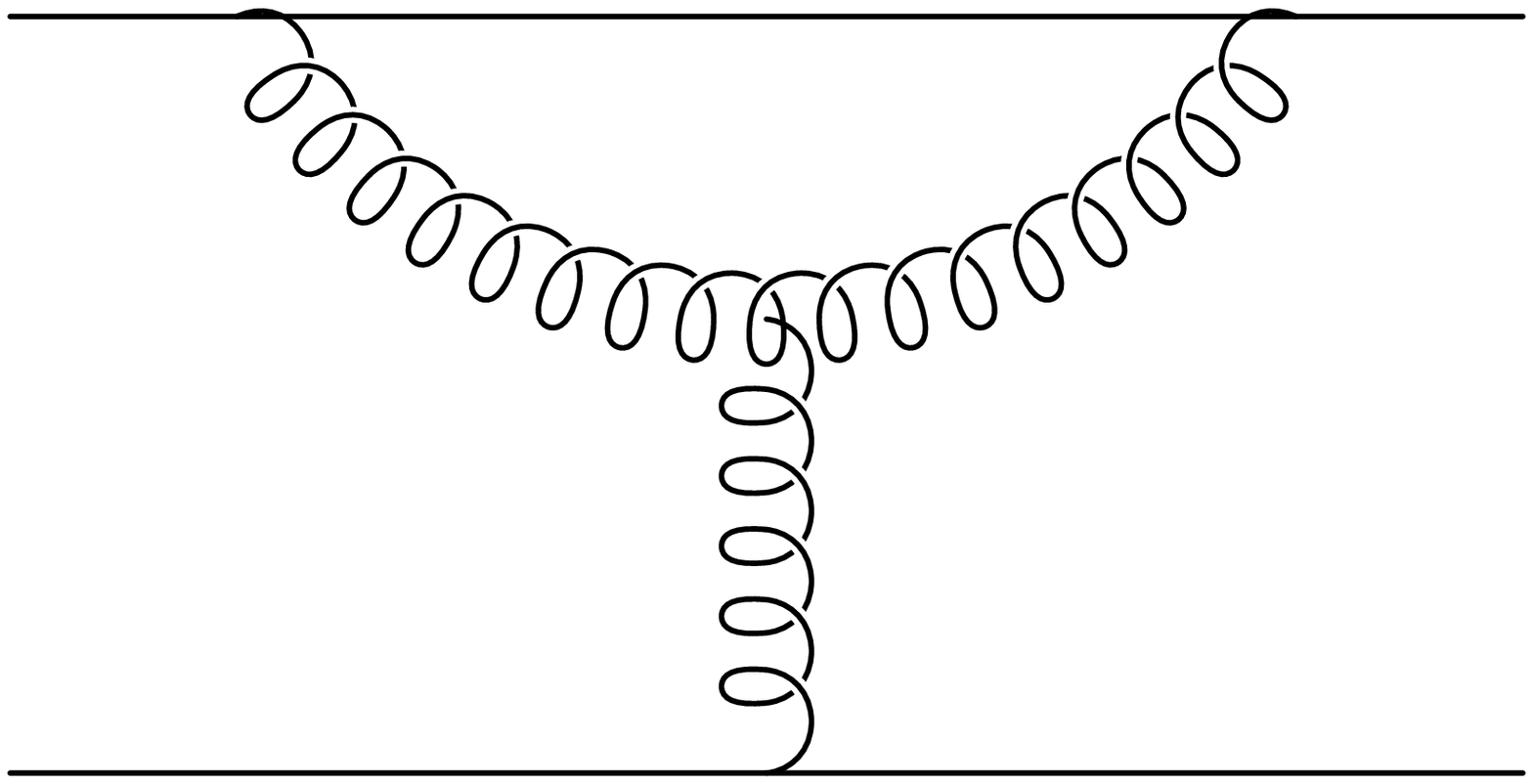}}  & C_A 
& {3 \over 2} 
\ln {\mu \over m}  & 0 & 0 & 0\\
&&&&&\\
\epsfxsize=2cm \lower20pt\hbox{\epsfbox{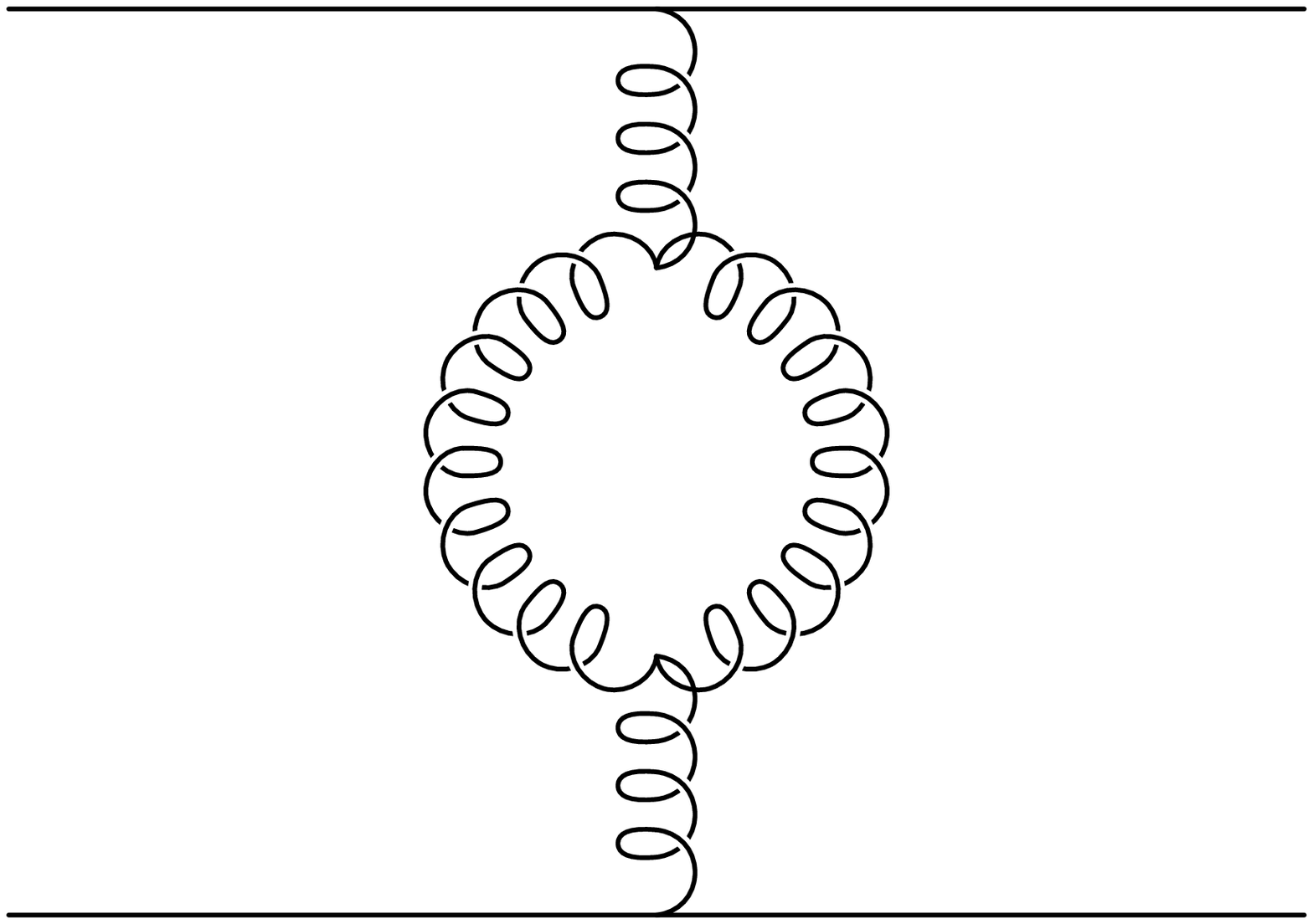}}  & C_A
& -{5 \over 6} 
\ln {k \over \mu} & -{5 \over 6} 
\ln {k \over \mu} & -{5 \over 6} 
\ln {k \over \mu} & 0 \\ &&&&&\\
\hline &&&&&\\
\epsfxsize=2cm \lower20pt\hbox{\epsfbox{9911_fd8.eps}} & C_A &  & 
& -{11 \over 6} \ln {k \over \mu} & \\ &&&&&\\
\epsfxsize=2cm \lower18pt\hbox{\epsfbox{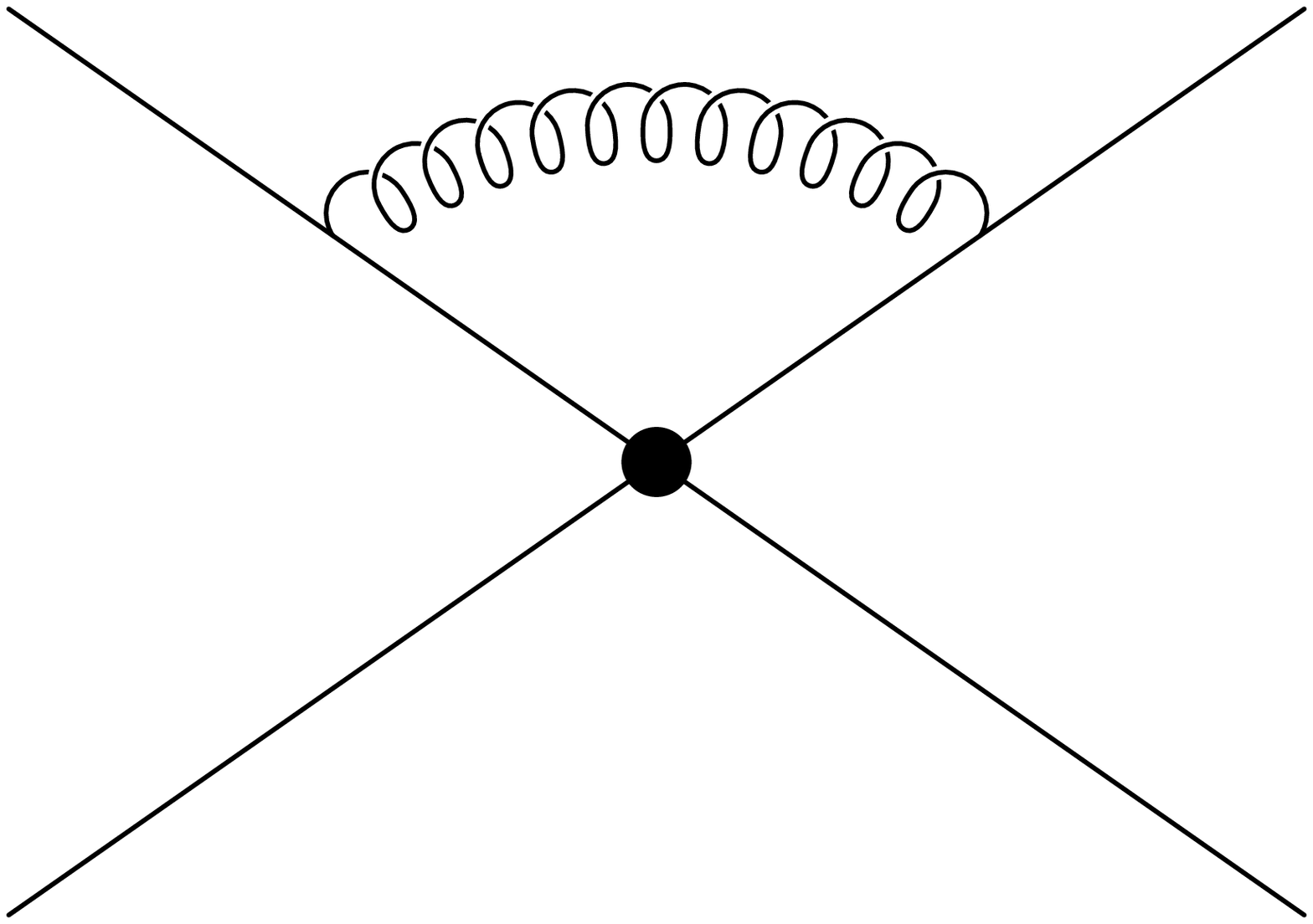}} & & & & & 0
\end{array}
\]
\end{table}

The sum of the QCD diagrams gives the net one-loop contribution to the static
potential (in the color singlet channel),
\begin{equation} \label{box:qcd}
V({\mathbf k})=-{4 \pi \alpha_s(\mu) C_F \over \abs{\mathbf k}^2} \left[1
-C_A{11\over 6} {\alpha_s (\mu) \over \pi} \ln {k \over \mu} \right],
\end{equation}
where only the logarithmic term has been retained. This gives the well-known
result that the potential can be rewritten as
\begin{equation}\label{ex:1}
V({\mathbf k})={4 \pi \alpha_s(\abs{\mathbf k}) 
\over \abs{\mathbf k}^2} T \cdot \bar T ,
\end{equation}
where $T \cdot \bar T=-C_F$ in the color singlet channel.

The box graph (the first diagram in Table~\ref{tab:box}) was computed using the
threshold expansion by Beneke and Smirnov~\cite{Beneke:1997}, and it is
interesting to see how the effective field theory reproduces the various
contributions. The hard part of the box graph is the matching condition between
QCD and NRQCD, i.e.\ a local two-fermion operator in the effective Lagrangian,
and is also equal to  the difference between the graphs computed in QCD and
HQET. The potential part of the box graph is reproduced in the effective theory
by the one-loop contribution from the iteration of two Coulomb
interactions~Fig.~\ref{fig:6}. The HQET value of the box diagram, $2 C_F \ln
\lambda/k$, is both infrared and ultraviolet finite if a gluon mass
$\lambda$ is introduced as an infrared regulator. This total contribution is
split, in the the threshold expansion, into an infrared divergent soft
contribution $2 C_F \ln \mu/k$, and an ultraviolet divergent ultrasoft
contribution $2 C_F \ln \lambda/\mu$. The sum of the soft and ultrasoft
contributions has no $1/\epsilon$ pole, since the ultraviolet and infrared
divergences cancel.\footnote{The cancellation is not really between an
ultraviolet and infrared divergence. In the effective theory, there are tadpole
graphs which are zero in dimensional regularization, and have the form of a
difference $1/\epsilon-1/\epsilon$ between an ultraviolet and infrared
divergence. One cannot characterize a $1/\epsilon$ pole as an infrared or
ultraviolet divergence if tadpole graphs are set to zero.}

The ultrasoft contribution vanishes in the full and effective theories, if the
infrared divergence is regulated by dimensional regularization. The non-trivial
contribution is the soft part of the box graph. The importance of this
contribution was pointed out by Griesshammer~\cite{Griesshammer:1998}, who
argued that one needed to introduce soft gauge fields, as well as soft quark
fields. In our approach, the soft part of the box graph is reproduced in the
effective theory by the loop graph shown in Fig.~\ref{fig:X5}, where the
interaction vertex is from the last two lines of Eq.~(\ref{nrqcd:1}). It is not
necessary to introduce soft quark fields, as advocated by Griesshammer. 

In QED, the soft part of the box and crossed-box graph cancel. This is
consistent with the effective Lagrangian Eq.~(\ref{nrqcd:1}), where the soft
interaction vertex vanishes for QED. In QED, the soft modes can be integrated
out directly at $\nu=1$, and replaced by local operators at the scale $\mu=m$.
This approach does not resum the logarithms of $v$ in the potential (which are
absent for QED), and so is not a satisfactory procedure for QCD.

The effective theory correctly reproduces the soft and ultrasoft contributions
to the static potential. The soft vertex in Fig.~\ref{fig:X5} is computed from the
Compton scattering  graphs in Fig.~\ref{fig:X3}. Fig.~\ref{fig:X5} reproduces
the sum of the soft part of the box, vertex and vacuum polarization corrections
to fermion-fermion scattering. We have seen that the soft vertex is
proportional to the commutator of two gauge fields, so the soft graph
Fig.~\ref{fig:X5} is proportional to $C_A$. This automatically implements the
cancellation between the various $C_F$ contributions to quark-antiquark
scattering in the full QCD calculation. An explicit computation of
Fig.~\ref{fig:X5} gives the contribution
\begin{equation}\label{eq:ir}
- {\left(4 \pi \alpha_s T\cdot \bar T \right) \over \abs{\mathbf k}^2}
{\alpha_s \over \pi}C_A \left(1 +{5 \over 6}\right) \ln {k \over \mus}
\end{equation}
to the scattering potential, where the first term (1) is an infrared divergent
contribution, and the second (5/6) is an ultraviolet divergent contribution.
Note that the soft graph is infrared divergent {\sl even if a gluon mass is
used as an infrared regulator}. The infrared divergent contribution of
Eq.~(\ref{eq:ir}) is converted to an ultraviolet divergent contribution if
tadpole graphs are included. Equation~(\ref{eq:ir}) agrees with the sum of the
soft contributions listed in Table~\ref{tab:box}.

\begin{figure}
\epsfxsize=8cm
\hfil\epsfbox{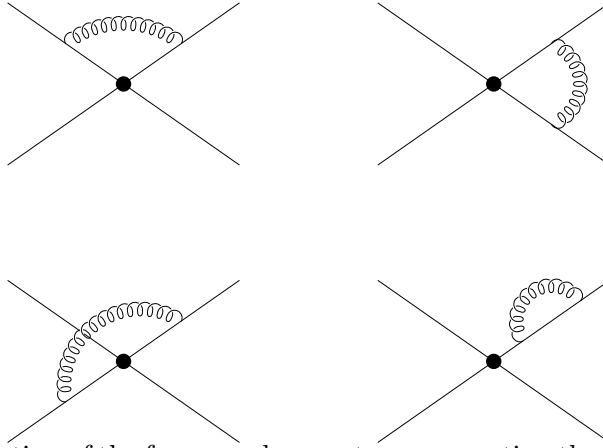}\hfill
\caption{Renormalization of the four-quark operator representing the
quark-antiquark potential by ultrasoft gluons.
\label{fig:usbox}}
\end{figure}

The ultrasoft contributions in the QCD theory add up to zero. The ultrasoft
contributions in the effective theory are the renormalization of the local
four-fermion quark-antiquark potential, shown in Fig.~\ref{fig:usbox}.
Each graph is ultraviolet divergent, and proportional to $\ln \muus/k$. The sum
of all the graphs is zero, in both the singlet and octet channels. In the
effective theory, ultrasoft radiative corrections do not renormalize the
quark-antiquark potential. They do, however, cause mixing between the leading
order potential, and corrections to the potential suppressed by powers of
$1/m$.

The quark-antiquark potential is VRG invariant. This implies that the coupling
$\alpha_s$ of the soft gluons must satisfy a $\beta$-function equation, where
the $\beta$-function  for the VRG is the same as the conventional one, since
$\mus=m\nu$, so that $\mus d/d\mus=\nu d/d\nu$. In other words, the
quark-antiquark potential takes the form Eq.~(\ref{box:qcd}), with the coupling
constant renormalized at the soft  scale $\mu=\mus=m\nu$. Choosing $m \nu =
\abs{\mathbf k}$ sums the leading logarithms, and gives Eq.~(\ref{ex:1}).

The effective theory has performed an interesting rearrangement of the terms in
the radiative correction to the static potential, compared with those in the
corresponding HQET computation in Feynman gauge. In the HQET computation, the
vertex and wavefunction renormalization graphs contribute $-C_A\ln k/\mu$, and
the vacuum polarization graph contributes an additional $-(5/6)C_A\ln k/\mu$.
The box diagram is finite, and does not contribute to the running of the static
potential. The vertex and wavefunction graphs involve ultrasoft loops, whereas
the vacuum polarization graph involves soft loops. In the effective theory, the
box graph contribution $\ln k/\lambda$ is broken up into a soft piece, $\ln
k/\mu$, and an ultrasoft piece, $\ln \mu/\lambda$. The ultrasoft piece cancels
the vertex and wavefunction graphs, leaving the soft contribution, so that the
static potential depends only on $\mus$, and not on $\muus$.\footnote{This is
true even when the running soft gluon coupling $\alpha_s(m\nu)$ and ultrasoft
gluon coupling $\alpha_s(m\nu^2)$ are used.} The breakup of the box graph into
soft and ultrasoft can be done without double counting in a mass-independent
scheme such as \ms~\cite{Beneke:1997}.

Terms in the one-fermion sector of the theory are renormalized by ultrasoft
gluons. The ultrasoft gluon coupling constant is renormalized due to their
self-interactions. All these graphs can be computed as for
HQET~\cite{Manohar:1997,Blok:1996,Bauer:1998}. The VRG anomalous dimension is
twice the usual anomalous dimension, since $\nu d/d\nu = 2 \muus d/d\muus$.

There are relations between the soft and ultrasoft gluon couplings in the
effective theory. The two couplings have a $\beta$-function that is related to
the QCD $\beta$-function, so that the soft coupling is $\alpha_{\rm
soft}(\nu)=\alpha_s(m\nu)$ and the ultrasoft coupling is $\alpha_{\rm
ultrasoft}(\nu)=\alpha_s(m\nu^2)$, where $\alpha_s(\mu)$ obeys the usual
renormalization group equation. This can be verified by explicit computation in
the effective theory.

\subsection{Integrating out a light fermion}

\begin{figure}
\epsfxsize=4cm
\hfil\epsfbox{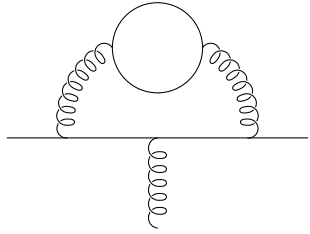}\hfill
\caption{A light fermion bubble contributions to an ultrasoft loop.
\label{fig:X15}}
\end{figure}

\begin{figure}
\epsfxsize=4cm
\hfil\epsfbox{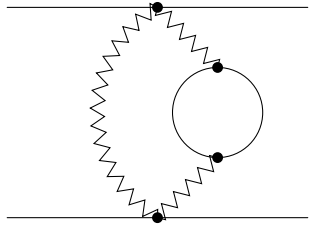}\hfill
\caption{A light fermion bubble contributions to a soft loop.
\label{fig:X16}}
\end{figure}

Consider the NRQCD theory with an additional fermion $\Psi$ of mass $m_\Psi$, 
with $\lqcd \ll m_\Psi \ll m$. At the scale $m$, one can match from QCD to an
effective theory that contains, in addition to the fields we have been
discussing, the fermion $\Psi$. At the scale $m$, the fermion $\Psi$ behaves
like a massless particle, so the effective theory contains soft and ultrasoft
fermion fields for $\Psi$, $\Psi_q(x)$ and $\Psi(x)$, respectively. The VRG
equation is used to scale the theory below $m$. At the velocity
$m\nu^2=m_\Psi$, i.e.\ $\nu=\sqrt{m_\Psi/m}$, the ultrasoft fermion modes
$\Psi(x)$ can be integrated out of ultrasoft loops such as Fig.~\ref{fig:X15},
and at the velocity $m\nu=m_\Psi$, i.e.\  $\nu={m_\Psi/m}$, the soft fermion
modes $\Psi_q(x)$ can be integrated out of soft loops such as
Fig.~\ref{fig:X16}.
This sums logarithms of $m/m_\Psi$ in both soft and ultrasoft loops.

\subsection{Two-loop running of the production current}\label{hoang}

A highly non-trivial check of the effective theory is the computation of the
two-loop running of the production current. This was computed by Hoang for
QED~\cite{Hoang:1997a,Hoang:1997b}, and the computation has been recently
extended to the non-Abelian case by Czarnecki and
Melnikov~\cite{Czarnecki:1998}, and by Beneke, Signer and
Smirnov~\cite{Beneke:1998}.

Consider production of a $\bar Q Q$ pair near threshold by a virtual photon.
The electromagnetic current in the full theory matches to the effective current
Eq.~(\ref{eff-1}) in the effective theory, where the two-loop matching
condition~\cite{Hoang:1997a,Hoang:1997b,Czarnecki:1998,Beneke:1998} between the
full and effective theories is given in Eq.~(\ref{eff-2}). The electromagnetic
current has no anomalous dimension in the full theory, which implies that in
the effective theory, $c_2$ must have an anomalous dimension,
\begin{equation}\label{anom2} \mu{d C(\mu)\over d\mu} =
-\pi^2  C_F \left({1\over3}C_F + {1\over2}C_A\right) \alpha_s^2
\end{equation}
at $\mu=m$.

The anomalous dimension Eq.~(\ref{anom2}) is of order $\alpha_s^2 v^0$, so we
need to compute all diagrams of this order in the effective theory. The
interactions needed are in the quark-antiquark potential to order $v^2$ in the
color singlet channel. The potential in the center of mass frame for the process
$Q({\mathbf p})+\bar Q({\mathbf -p}) \to Q({\mathbf p^\prime})+\bar 
Q({\mathbf -p^\prime})$ with momentum transfer $\mathbf k=p-p^\prime$ is
(borrowing the notation of Titard and Yndurain~\cite{Titard:1994,Titard:1995})
\begin{eqnarray}
V = V_C \left( {1 \over \abs{\mathbf k}^2 } + {\abs{\mathbf
p}^2 \over m^2 \abs{\mathbf k}^2}  \right) + V_{\abs{\mathbf k}}
{\pi^2 \over m  \abs{\mathbf k}} +{V_2\over m^2} + {V_{hf}\over m^2} {\mathbf S}^2 + 
{ V_{LS} \over m^2} \Lambda({\mathbf k}) + {V_T \over m^2} T
\end{eqnarray}
where
\begin{eqnarray}
{\mathbf S}&=&{\bsigma_1 + \bsigma_2 \over 2},\\
\Lambda({\mathbf k}) &=& -i {\mathbf S \cdot (k\times p) \over k^2},\\
T&=&\bsigma_1\cdot \bsigma_2 -{3\over {\mathbf k}^2} \mathbf(k \cdot
\bsigma_1)(k\cdot \bsigma_2)
\end{eqnarray}
and the coefficients we need are
\begin{eqnarray}
V_C(\mu=m)&=&-{4 \pi \alpha_s(m) C_F},\nonumber \\
V_{\abs{\mathbf k}}(\mu=m) &=& \alpha_s^2(m) C_F \left({1\over 2} C_F - C_A
\right), \nonumber\\
V_{2}(\mu=m) &=& 0,\nonumber \\
V_{hf}(\mu=m) &=& {4 \pi \alpha_s(m) C_F \over 3} ,\label{ex:coef}
\end{eqnarray}
using tree-level matching at $\mu=m$ for $V_{C,2,hf}$ and one-loop matching at
$\mu=m$ for $V_{\abs{\mathbf k}}$. Note that the $1 /\abs{\mathbf k}^2$ and
$\abs{\mathbf p}^2 /\abs{\mathbf k}^2$ terms have the same coefficient, which
follows from reparameterization invariance. In addition to the above terms, we
also need the $p^4/8m^3$ kinetic energy correction in the Lagrangian, whose
coefficient is also fixed by reparameterization invariance.

\begin{figure} 
\epsfxsize=10cm 
\hfil\epsfbox{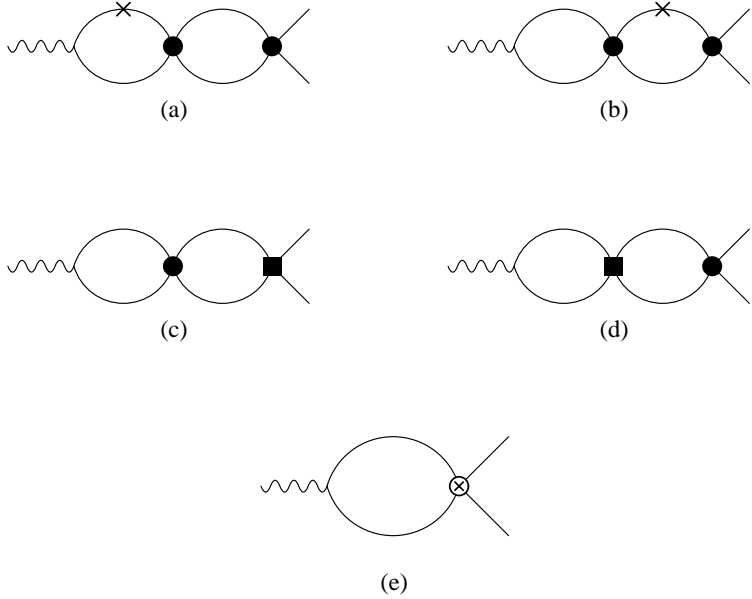}\hfill
\caption{Diagrams which contribute to the two-loop anomalous dimension of the
production current. The dot and box represent terms in the tree-level potential
at order 1 ($V_C$), and $v^2$ ($V_C$, $V_2$, $V_{hf}$, $V_{LS}$, $V_T$) 
respectively.
The $\otimes$ is an insertion of the order $v^2$ term proportional to
$\alpha_s^2/m \abs{\mathbf k}$ in the one-loop potential ($V_{\abs{\mathbf
k}}$). The $\times$ is an insertion of the $p^4/8m^3$ kinetic energy
correction. \label{fig:2loop}}
\end{figure}

The diagrams which contribute to the two-loop anomalous dimension of the
production current in the effective theory are shown in Fig.~\ref{fig:2loop}.
The contribution from the diagrams to the anomalous dimension will be called
$\gamma_a$--$\gamma_e$, respectively, where
\begin{equation}
\mus{d C \over d\mus} = \nu {d C \over d\nu}=\gamma.
\end{equation}
One finds that
\begin{eqnarray}
\gamma_a & =& 0,\nonumber\\
\gamma_b & =& -{1\over 64 } V_C^2,\nonumber\\
\gamma_c & =& -{1\over 16}  V_C \left(V_2 + 2 V_{hf} \right) 
-{1\over 32 }V_C^2,\nonumber\\
\gamma_d & =& -{1\over 32 }V_C^2,\nonumber\\
\gamma_e & =& {\pi^2 \over 2} V_{\abs{\mathbf k}}
\end{eqnarray}
and the total anomalous dimension is
\begin{eqnarray}
\gamma&=&\gamma_a+\gamma_b+\gamma_c+\gamma_d+\gamma_e=
-{5 V_C^2 \over 64 } - {V_C \left(V_2 + 2 V_{hf} \right) \over 16} + 
{ \pi^2 V_{\abs{\mathbf k}} \over 2}
\end{eqnarray}
which agrees with the known result for QCD at $\mu=m$,  Eq.~(\ref{anom2}), when
the coefficients Eq.~(\ref{ex:coef}) are used.

To complete the renormalization group analysis of the two-loop anomalous
dimension, one needs the running values for $V_C$, $V_{hf}$ and
$V_{\abs{\mathbf k}}$, which will be presented elsewhere.  Note, however,
that our analysis disagrees with the results of \cite{Beneke:1999}, in which
the running of these terms in the potential was neglected.

\section{Conclusions}

We have discussed a formulation of nonrelativistic QCD that can be used with a
mass-independent subtraction scheme such as \ms. The effective theory passes
several non-trivial checks: (1) it has a consistent $v$ power counting
expansion, (2) it correctly reproduces the running of the quark-antiquark
potential at one loop, and (3) it correctly reproduces the two-loop running of
the production current. The effective theory allows one to simultaneously treat
the momentum regions of order $mv$ and $mv^2$.

In the way we have formulated the effective theory, all large logarithms have
been summed at the scale $\nu=v$. At this point, one can integrate out the soft
and modes, and at the same time switch from a theory of quarks and antiquarks
to a theory of quarkonia, i.e.\ to an effective Lagrangian representing
quarkonia interacting with background color fields, as first studied by
Voloshin~\cite{Voloshin:1979} and Leutwyler~\cite{Leutwyler:1981}.

The methods described here should also be applicable to other nonrelativistic
field theories, such as those describing nucleon-nucleon scattering at low
energies~\cite{Seki:1998}.

\acknowledgments We would like to acknowledge helpful discussions with J.~Kuti,
C.~Morningstar and M.B.~Wise, and would like to  thank A.~Czarnecki for
provided details about his computation of the two-loop anomalous dimension, and
F.~Yndurain for correspondence about the one-loop corrections to the
quark-antiquark potential. We are particularly indebted to A.H.~Hoang, for
numerous discussions on the two-loop anomalous dimensions, and for checking
some of our calculations.

This work was supported in part by Department of Energy grants
DOE-FG03-97ER40546 and DOE-ER-40682-137.

\end{document}